\documentclass[aps,prstab,twocolumn,showkeys,groupedaddress]{revtex4-2}

\usepackage{graphicx}%
\usepackage{multirow}%
\usepackage{amsmath,amssymb,amsfonts}%
\usepackage{amsthm}%
\usepackage{mathrsfs}%

\usepackage[T1]{fontenc}
\usepackage{textcomp}
\usepackage{lmodern}
\usepackage [latin1]{inputenc}

\begin{document}


\title[A two-color dual-oscillator infrared free-electron laser]{A two-color dual-oscillator infrared free-electron laser}



\author{Wieland Sch{\"o}llkopf}
\email{wschoell@fhi-berlin.mpg.de}
\author{Sandy Gewinner}
\author{Marco {De}  Pas}
\author{Heinz Junkes}
\author{Sebastian Kray}
\author{William Kirstaedter}
\author{{William B. Colson}}
\email[Consultant to the Fritz-Haber-Institut der Max-Planck-Gesellschaft, Faradayweg 4-6, 14195 Berlin, Germany]{}
\author{{David H. Dowell}}
\email[Consultant to the Fritz-Haber-Institut der Max-Planck-Gesellschaft, Faradayweg 4-6, 14195 Berlin, Germany]{}
\author{{Stephen C. Gottschalk}}
\email[Consultant to the Fritz-Haber-Institut der Max-Planck-Gesellschaft, Faradayweg 4-6, 14195 Berlin, Germany]{}
\author{{John W. Rathke}}
\email[Consultant to the Fritz-Haber-Institut der Max-Planck-Gesellschaft, Faradayweg 4-6, 14195 Berlin, Germany]{}
\author{{Tom J. Schultheiss}}
\email[Consultant to the Fritz-Haber-Institut der Max-Planck-Gesellschaft, Faradayweg 4-6, 14195 Berlin, Germany]{}
\author{{Alan M. M. Todd}}
\email[Consultant to the Fritz-Haber-Institut der Max-Planck-Gesellschaft, Faradayweg 4-6, 14195 Berlin, Germany]{}
\author{{Lloyd M. Young}}
\email[Consultant to the Fritz-Haber-Institut der Max-Planck-Gesellschaft, Faradayweg 4-6, 14195 Berlin, Germany]{}
\author{Akash Chandra Behera}
\author{Am{\'e}rica Y. Torres-Boy}
\author{Martin Wolf}
\author{Alexander Paarmann}
\author{Gert {von}  Helden}
\author{Gerard Meijer}

\affiliation{Fritz-Haber-Institut der Max-Planck-Ge\-sell\-schaft, Faradayweg 4-6, 14195 Berlin, Germany}



\date{\today}

\begin{abstract}
We report on the design and performance of a two-color dual-oscillator infrared free-electron laser (FEL). The mid-infrared (MIR) FEL at the Fritz Haber Institute (FHI FEL) has been upgraded to include a second oscillator FEL beamline that permits lasing in the far-infrared (FIR) regime from 4.5 $\mu$m to 175 $\mu$m. In addition, a 500 MHz kicker cavity has been installed downstream of the electron accelerator. It allows to deflect electron bunches of up to 50 MeV energy alternately left and right by an angle of $\pm 2^\circ$. It can, thus, split the high-repetition-rate (1 GHz) electron bunch train from the accelerator into two bunch trains of 500 MHz repetition rate each; one is steered to the MIR FEL and the other one to the new FIR FEL. In this two-color mode of simultaneous, synchronized operation the wavelengths in both FELs can be tuned independently over wide ranges of up to a factor of four each by undulator-gap variation. In addition, two-color operation is also available at reduced repetition rates (e.g.\ 55.6 MHz of both MIR and FIR pulses), as needed for some applications. This unique two-color mode opens up a wealth of novel user applications such as, MIR-FIR pump-probe experiments.
\end{abstract}


\maketitle

\section{The FHI FEL Facility}

The first above-threshold operation of a free-electron laser (FEL) oscillator was demonstrated almost fifty years ago, at Stanford University, generating light in the infrared, near 3.4 $\mu$m \cite{DEACON:1977ab}. Later, two of the leading scientists on that project elaborated on the possibility of setting up a two-color infrared free-electron laser, in which two synchronous optical beams are produced at different frequencies \cite{SCHWETTMAN:1989aa}. The application of such {``}{\it more useful [...] than the conventional one-color FEL}'' system for condensed matter molecular dynamics was extensively discussed \cite{Dlott:1989}. However, two-color operation of a dual-oscillator FEL, based on the high-frequency deflection of electron bunches, exactly as proposed in 1989 \cite{SCHWETTMAN:1989aa} and as presented here, has - to the best of our knowledge - never been implemented before. 

To realize such a two-color dual-oscillator scheme we have upgraded the infrared FEL at the Fritz Haber Institute (FHI FEL) which provides intense, pulsed mid-infrared (MIR) radiation continuously tunable in the wavelength range from 2.9 $\mu$m to 50 $\mu$m \cite{SPIE_2015}. Operational for in-house user experiments since 2013, this MIR FEL has been used in diverse experiments ranging from spectroscopy of clusters and (bio-)molecules in the gas-phase to nonlinear solid-state spectroscopy and surface science resulting in more than 110 peer-reviewed publications so far.

The upgrade of the FHI FEL includes installation of a second oscillator FEL beamline, as outlined in Fig.\ \ref{fig:overview}, which permits lasing in the far-infrared (FIR) regime from 4.5 $\mu$m to 175 $\mu$m \cite{FEL_2019,FEL_2022,IPAC2024,FEL_2024}. In addition, a 500 MHz kicker cavity has been installed downstream of the electron accelerator. It operates in a dipole mode using the strong electric field between two vanes to deflect electron bunches up to 50 MeV energy alternately left and right by an angle of $\pm 2^\circ$. It can split the high-repetition-rate (1 GHz) electron bunch train arriving from the accelerator into two bunch trains of 500 MHz repetition rate each; one is steered to the MIR FEL and the other one to the FIR FEL. This way, the kicker cavity enables simultaneous operation of the MIR and FIR FELs in a highly synchronized way. In this two-color mode the wavelengths in both FELs can be tuned independently by undulator gap variation over wide ranges of up to a factor of four each. In 2025 we started routine user operation of the two-color mode giving our user groups the means to perform novel experiments not possible at any other facility, including, for instance, 
double-resonance, two-color pump-probe, or 2D-IR spectroscopy.

In this paper we describe the design and performance of the FHI two-color dual-oscillator infrared free-electron laser in detail. The paper is organized as follows. In Section \ref{sec:MIR} we briefly review the specifications of the accelerator and the MIR FEL before we describe the upgrade design including the FIR FEL and the kicker cavity in Section \ref{sec:2color:upgrade} in detail. Subsequently, we present the performance of the FIR FEL (Section \ref{sec:FIR}) and the two-color mode (Section \ref{sec:2color:performance}). We provide a summary and an outlook on planned two-color user experiments in the Conclusions Section.

\section{Electron accelerator and MIR FEL} 
\label{sec:MIR}

\begin{figure}
 \includegraphics[width=0.48\textwidth]{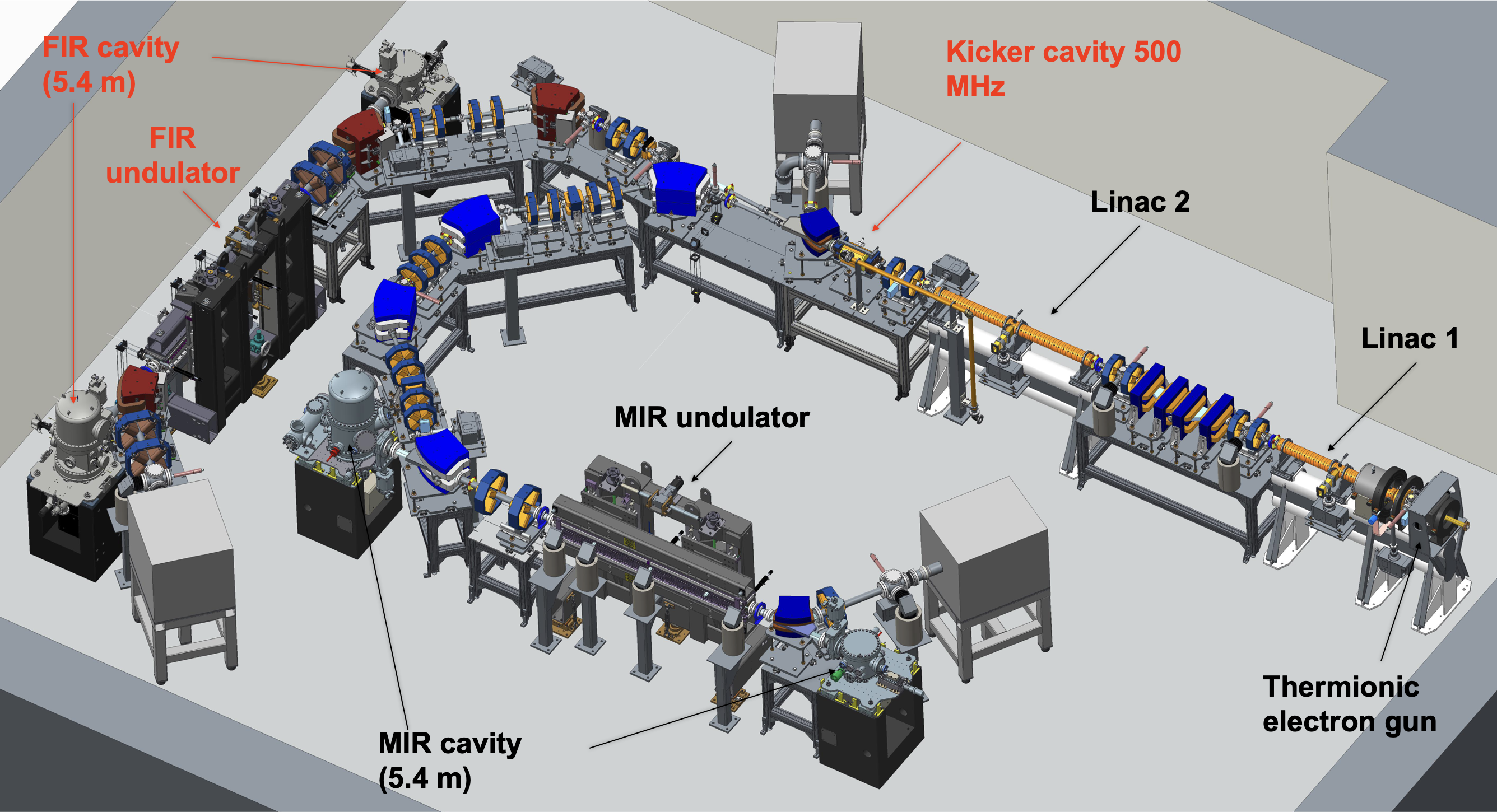}
\caption{Layout of the FHI FEL. Red labels indicate the new components installed as part of the two-color upgrade commissioned in 2023, whereas black labels indicate the parts that have been in operation since 2013.} \label{fig:overview}
 \end{figure}

The electron accelerator system is based on a gridded thermionic DC electron gun and a sub-harmonic (1 GHz) buncher cavity. Thereafter, a normal-conducting accelerator including two standing-wave S-band linac stages delivers bunches ($> 200$ pC charge, few ps long) at energies from 15 to 50 MeV with low longitudinal ($< 50$ keV\,ps) and transverse ($< 20 \pi$ mm\,mrad) emittance. The temporal structure is characterized by a bunch repetition rate of 1 GHz (with optional reduced repetition rates) with a burst (macro-bunch) length of typically 10 $\mu$s at 10 Hz macro-bunch repetition rate. When operating the MIR FEL, the electron beam is delivered by two $90^{\circ}$ isochronous achromats to the MIR FEL, as can be seen in Fig.\ \ref{fig:overview}. 

The MIR pulses are generated by the electrons passing through a planar hybrid-magnet undulator located within an IR cavity of $L_0 = 5.4$ m nominal length. The 2.2-m-long wedged-pole undulator includes 50 periods, each of 40 mm length \cite{FEL_2012c}. At a minimum gap of 17.4 mm, it reaches a root-mean-square undulator parameter $K_{\rm rms} = 1.51$. The design and performance parameters of the MIR undulator are listed and compared to the new FIR undulator in Table\ \ref{tab:U68} below. A quadrupole doublet upstream of the MIR undulator allows for matching the electron bunches into the vertically focussing undulator. To free up space needed for installing the quad doublet the MIR undulator position is shifted downstream by 50 cm with respect to the IR cavity center. The optical mode of the spherical concave cavity mirrors is characterized by a waist at the undulator center and a Rayleigh range of 1 m corresponding to about half the undulator length.

\begin{figure}
 \includegraphics[width=0.48\textwidth]{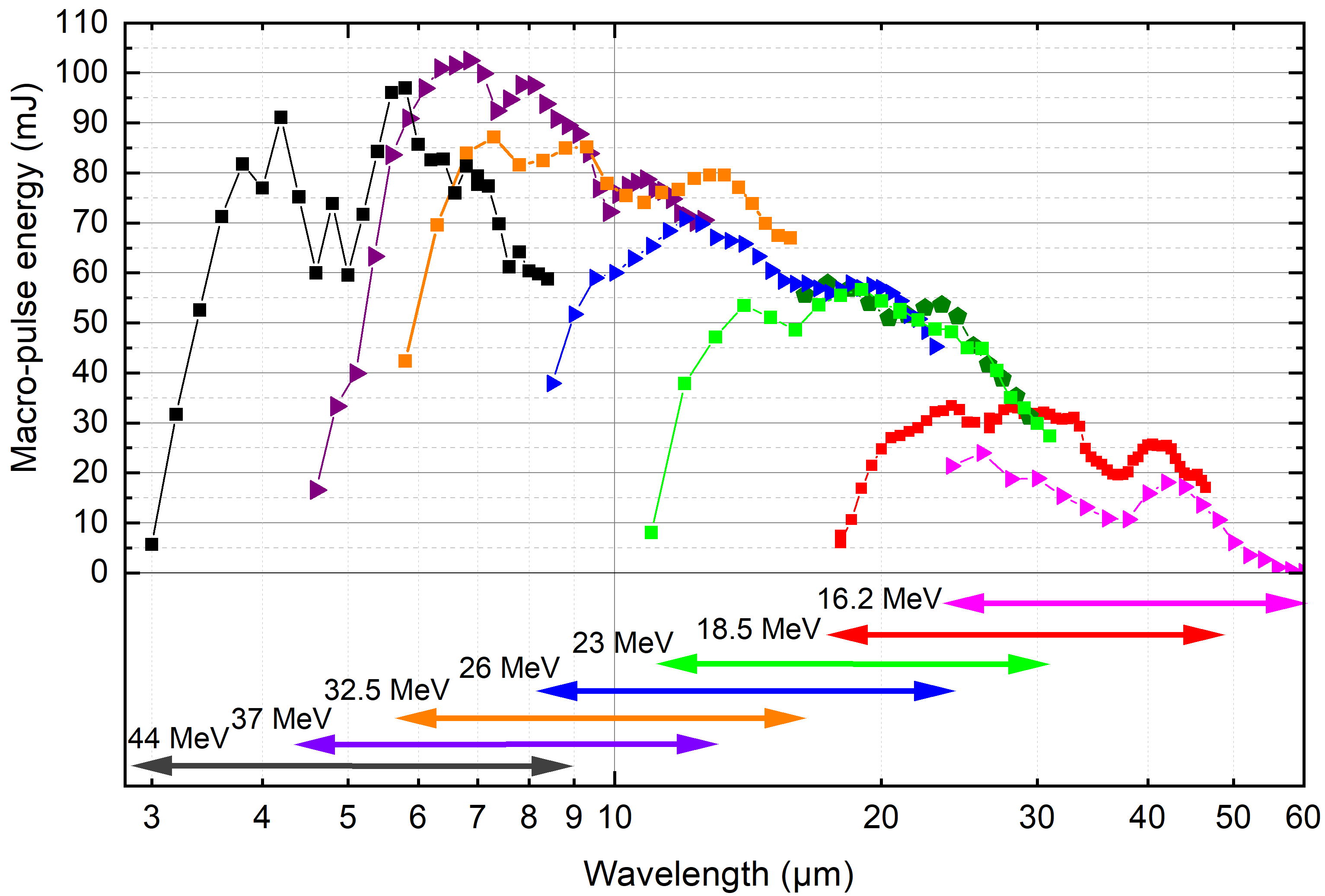}
\caption{Macro-pulse energy of the MIR FEL which has been in user operation since 2013. Each of the seven curves is measured by scanning the undulator gap for the given electron energy indicated next to the colored arrow at the bottom. All data were taken at 1 GHz pulse repetition rate and for narrow spectral widths (full width at half maximum of the FEL spectrum of 0.5\% or less relative to the central wavelength).} \label{fig:MIRpwr}
 \end{figure}

Measurements of the FEL macro-pulse energy as a function of wavelength $\lambda$ are shown in Fig.\ \ref{fig:MIRpwr}. Each trace in the figure corresponds to an individual accelerator tune-up at the indicated electron energy. The typical energy of $\sim$10 $\mu$J per micro pulse corresponds to $\sim$100 mJ per macro pulse. The central wavelength of the MIR FEL can be set to any value between 2.9 and 60 $\mu$m, thereby covering the full mid-infrared range \cite{SPIE_2015}. The micro-pulse length and, hence, the spectral width of the radiation can be varied by fine adjustment of the FEL cavity length to $L_0 - \Delta L$, where $\Delta L = p \lambda$ is referred to as the cavity detuning which is typically set in proportion to the wavelength $\lambda$ with $p = 0.5 \dots 10$. For most spectroscopic applications the spectral width is set to less than 0.5\% (full width at half maximum) of the central wavelength by choosing a relatively large cavity detuning of  $\Delta L = 3.5 \dots 8 \lambda$. This corresponds to a Fourier-limited micro-pulse length of a few ps. Ultrashort pulses as short as 0.5 ps of high peak power, as sometimes needed for non-linear spectroscopy or time-resolved measurements, can also be generated at a correspondingly broader spectrum of several percent by choosing a small detuning $\Delta L = 0.5 \dots 2 \lambda$ \cite{Kiessling_2018}.

A 1 GHz RF pulse ($\sim$200 W in-pulse power) applied to the grid of the DC gun leads to a 1 GHz electron-bunch repetition rate. In this standard mode of operation there are 36 infrared pulses, equally spaced by 30 cm, oscillating back and forth in the FEL cavity. In addition, the cavity length of 5.4 m permits operation of the FEL at the reduced electron-bunch repetition rates listed in Table\ \ref{tab:rates}. For instance, at an electron bunch repetition rate of 55.6 MHz two optical pulses are generated in either the MIR or the FIR FEL. This mode is needed for user experiments where the FEL pulses are synchronized with pulses from a 55.6 MHz tabletop laser. Furthermore, as will be discussed below, operating the electron gun at 111.1 MHz delivers 55.6 MHz repetition rate operation in both FELs in two-color lasing mode (two optical pulses in each FEL cavity). Operation at a reduced repetition rate mode has been implemented by applying short ($\sim$1 ns) high-voltage pulses ($\sim$160 V peak), generated by an in-house-made broadband solid-state pulsed amplifier at the respective repetition rate, to the grid of the gun. The electron bunch repetition rate modes printed in bold in Table\ \ref{tab:rates} are regularly requested in user applications. 


\begin{table}[hbt]
\small
  \caption{Feasible repetition rates of the electron micro-bunches, numbers of IR pulses simultaneously circulating the optical cavity, and corresponding pulse-to-pulse time separations that are consistent with a 5.4 m long optical cavity (36 ns round-trip time). Besides the standard 1 GHz mode FHI FEL users regularly request 55.6 MHz operation (2-pulse mode). In addition, when operating in two-color mode, the electron gun and accelerator are operated at either 1 GHz, resulting in a 500 MHz mode operation of each MIR and FIR FEL, or at 111.1 MHz resulting in a 55.6 MHz operation (2-pulse mode) in each MIR and FIR FEL.}
  \label{tab:rates}
  \begin{tabular*}{0.3\textwidth}{@{\extracolsep{\fill}}|r|r|r|}
    \hline
     Bunch           & No.\ of & Pulse to  \\
     repetition      & pulses in & pulse   \\ 
      rate             &  cavity   &  separation \\
      \hline
     \bf{1 GHz} &   \bf{36}  &    \bf{1 ns}      \\        
      500 MHz  &         18  &         2 ns  \\
      333.3 MHz &       12  &        3 ns      \\
      250 MHz  &           9  &         4 ns         \\  
      166.7 MHz  &        6 &         6 ns      \\
      \bf{111.1 MHz} &  \bf{4} &  \bf{9 ns}    \\ 
      83.3 MHz  &          3  &        12 ns    \\
      \bf{55.6 MHz} &  \bf{2}  &    \bf{18 ns} \\   
      27.8 MHz &           1   &        36 ns \\
      \hline
\end{tabular*}
\end{table}

\section{Two-Color Upgrade of FHI FEL}
 \label{sec:2color:upgrade}
 
The design of the two-color FHI FEL upgrade \cite{FEL_2019,FEL_2022,IPAC2024} presented here was based on three user requirements: (i) extension of the wavelength range of the facility into the FIR/THz regime out to a wavelength of at least 150 $\mu$m (2 THz) by setting up a second oscillator FEL; (ii) ability to operate both MIR and FIR FELs simultaneously, thereby offering users two-color capabilities for pump-probe experiments; and (iii) design the new FIR FEL such that the wavelength ranges (undulator-gap tuning ranges) of both FELs overlap at any given electron energy.

The upgraded FHI FEL layout fulfilling these requirements is shown in Fig.\ \ref{fig:overview}, where the new FIR FEL beamline is visible on the left. A 500 MHz side-deflecting-cavity (kicker cavity) was installed downstream of the accelerator. When this cavity is powered, it deflects the electron bunches alternately left and right into the MIR and FIR beamlines, respectively. Using a side-deflecting RF cavity that operates at one-half of the bunch repetition rate to split a high-repetition-rate bunch train in two was first proposed by Schwettman and Smith in 1989 \cite{SCHWETTMAN:1989aa}. However, to the best of our knowledge, such a scheme was not implemented anywhere in the context of an infrared FEL until now. 
 
In the following subsections we describe the design of the kicker cavity, the electron beam transfer to the FIR FEL as well as the FIR optical cavity and undulator in detail.

\subsection{500 MHz Kicker Cavity}
\label{sec:kicker}

\begin{figure}
 \includegraphics[width=0.42\textwidth]{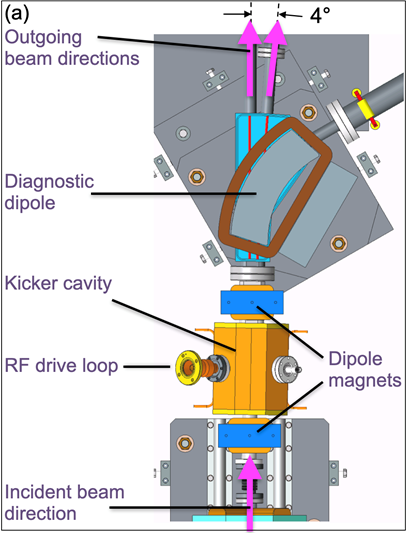}
 \includegraphics[width=0.5\textwidth]{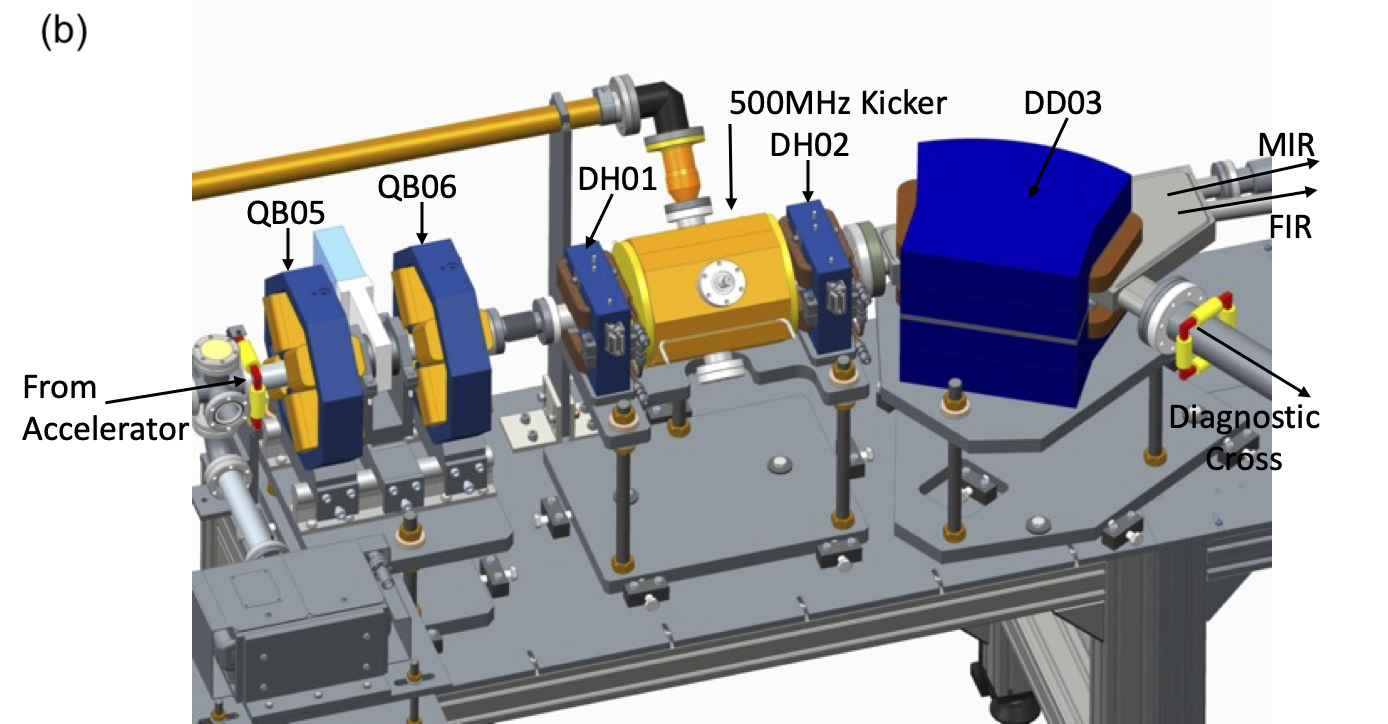}
\caption{Top view (a) and side view (b) of the kicker-cavity beamline section. Electron bunches coming from the accelerator can be alternately separated by $4^{\circ}$ into the MIR (straight-on direction) and FIR ($4^{\circ}$ to the right) beamlines. The two auxiliary dipole magnets labeled DH01 before and DH02 after the kicker cavity are indicated. In two-color operation each of them deflects the electrons by $1^{\circ}$ to the right. In combination with the $\pm 2^{\circ}$ deflection by the kicker cavity this results in a total deflection of zero and of $4^{\circ}$ to the right for electron bunches steered to the MIR and FIR FEL branches, respectively. As barely visible in (a), the axis of the kicker cavity is tilted by $1^{\circ}$ to the right with respect to the accelerator axis. The $1^{\circ}$ deflection by DH01 in two-color operation thus ensures that the electrons enter the kicker cavity collinear with its axis.} \label{fig:kicker}
 \end{figure}

The layout of the kicker beamline section is shown in Fig.\ \ref{fig:kicker}. A strong 500 MHz RF field applied to the kicker cavity generates a transverse horizontal electric field (peak field of up to 11.5 MV/m). It deflects electrons of up to 50 MeV maximum energy sideways by $\pm 2^{\circ}$. This allows every second electron micro-bunch to be sent to the MIR FEL and every other second bunch to the FIR FEL branch after a relatively short drift region behind the cavity, as shown in Fig.\ \ref{fig:kicker}. The kicker-cavity setup includes two auxiliary dipole magnets before and behind the cavity labelled DH01 and DH02, respectively, in Fig.\ \ref{fig:kicker}(b). This configuration makes three different modes of operation possible: (i) in the MIR FEL mode both dipoles and the RF to the kicker cavity are off, 100\% of the electron beam goes to the MIR FEL; (ii) in the FIR FEL mode each dipole deflects the electron beam by $2^{\circ}$ to the right while the RF to the kicker cavity is off, 100\% of the beam goes to the FIR FEL; and (iii) in the two-color mode each of the two dipole magnets is set to deflect the beam by $1^{\circ}$ to the right while the kicker cavity is energized to deflect by $\pm 2^{\circ}$. This way, in the two-color mode, the 1 GHz repetition-rate bunch train coming from the accelerator is split into two 500 MHz repetition-rate bunch trains feeding both MIR and FIR FEL.

The $60^{\circ}$-deflection diagnostic dipole labelled DD03 in Fig.\ \ref{fig:kicker}(b) is not energized in any of these modes. With the kicker and small dipoles off, the diagnostic dipole can be used to deflect the electron beam into a third (diagnostic) beamline for energy-spread measurements.

\begin{figure}
 \includegraphics[width=0.485\textwidth]{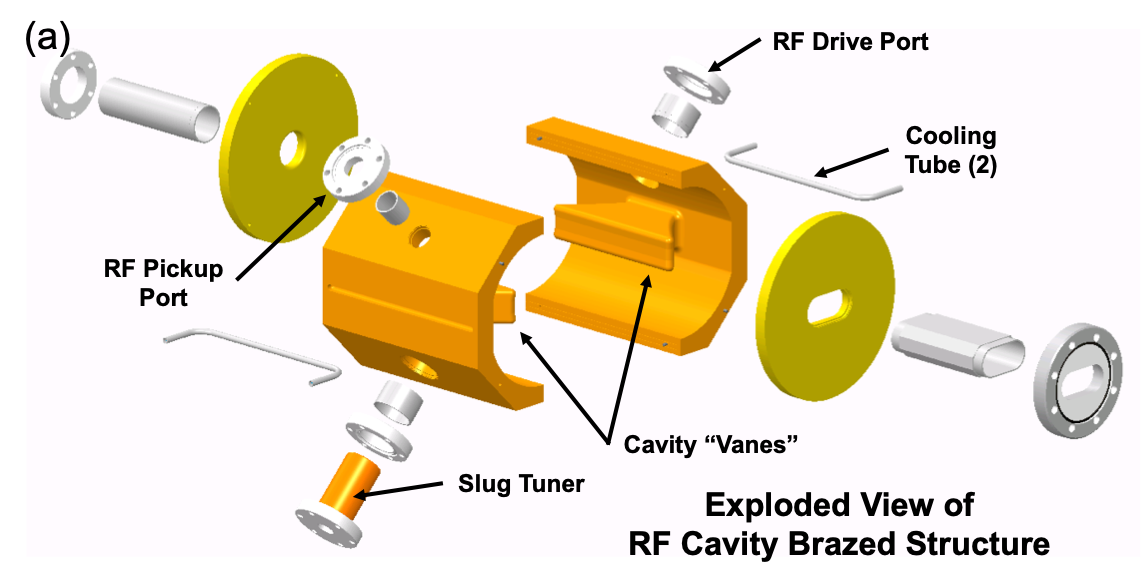}
  \includegraphics[width=0.225\textwidth]{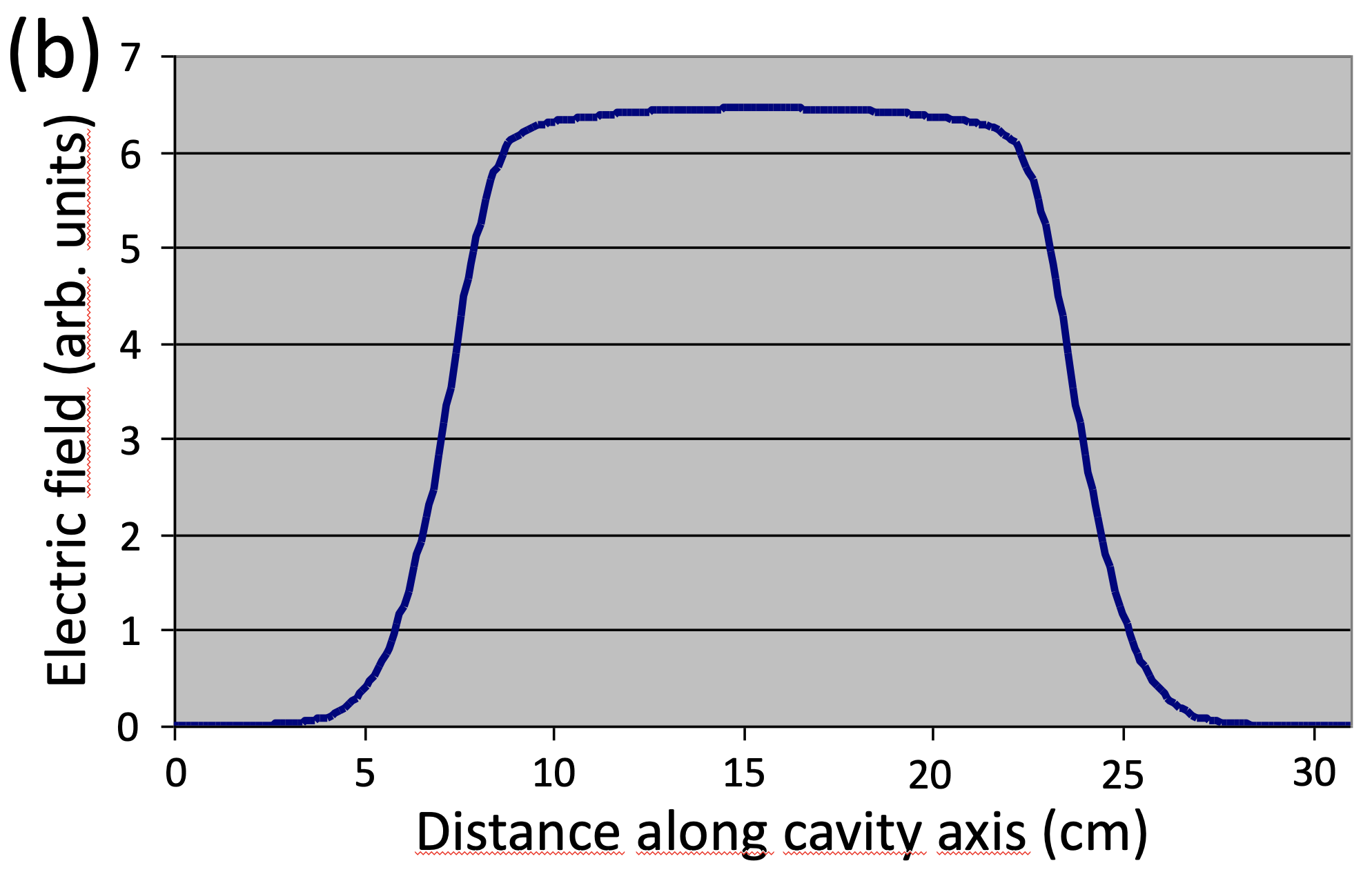}
 \includegraphics[width=0.25\textwidth]{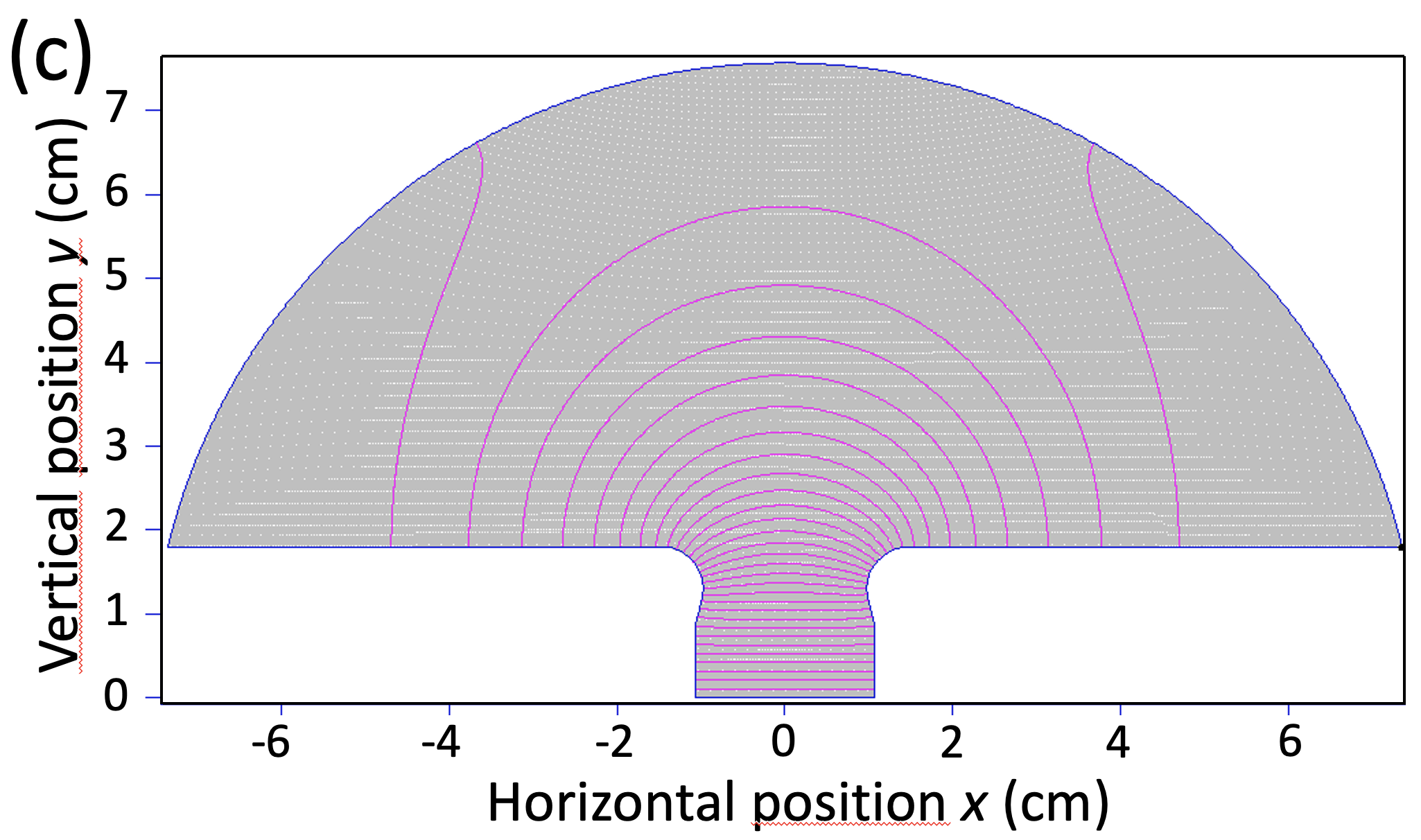}
 \includegraphics[width=0.48\textwidth]{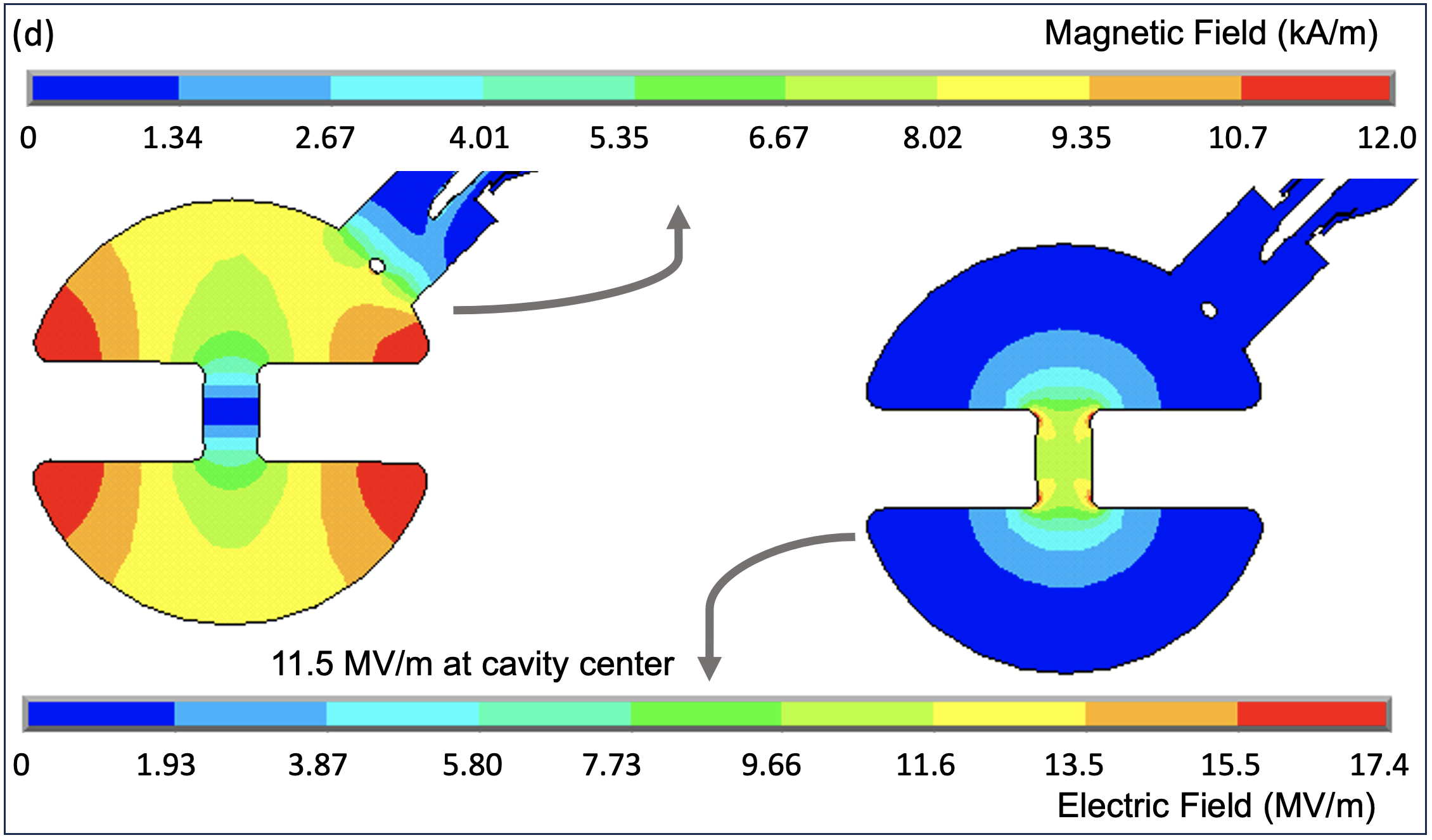}
 \includegraphics[width=0.48\textwidth]{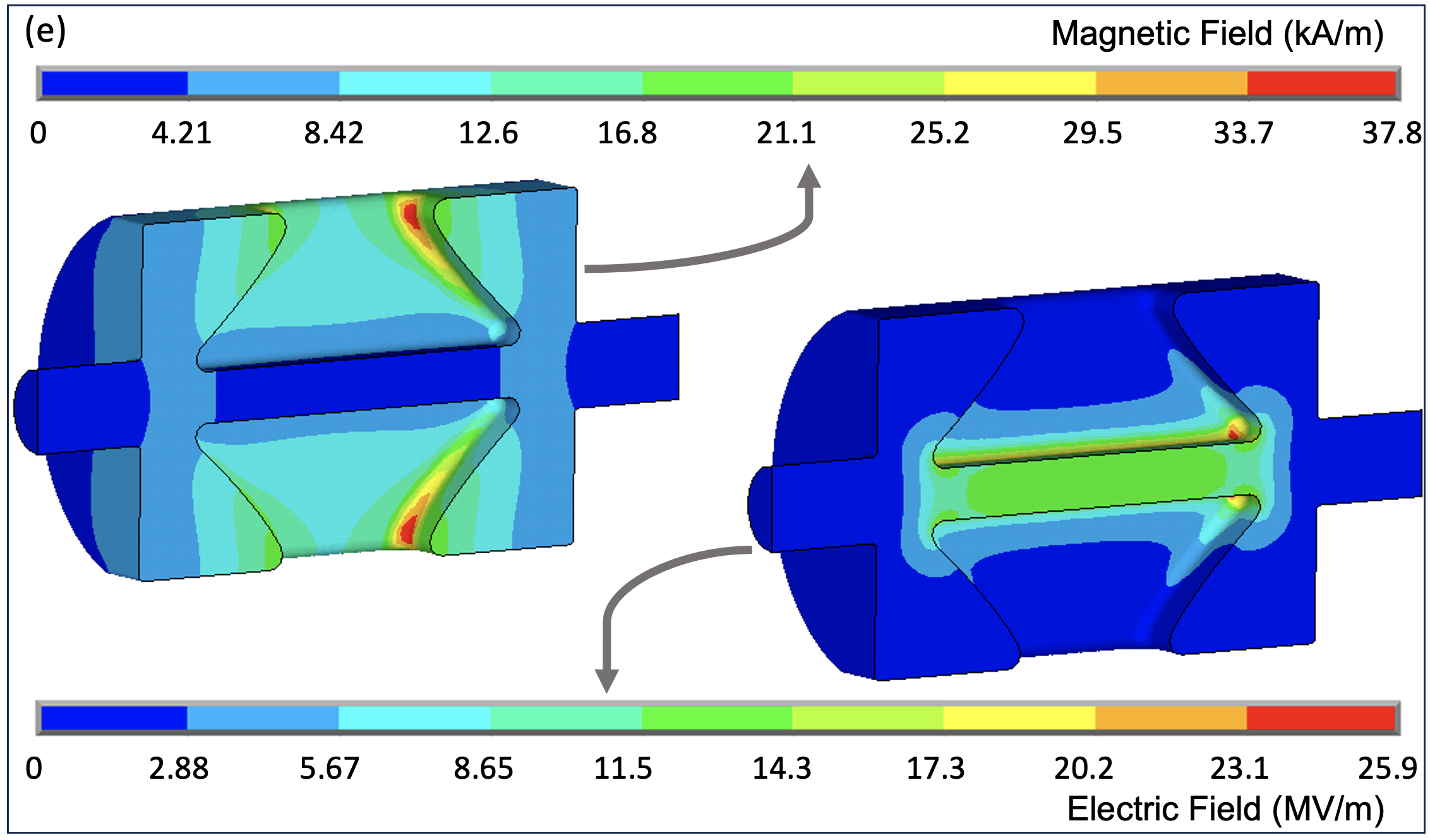}
\caption{(a) Internal shape of the kicker cavity where a 500 MHz transverse RF electric field builds up between two copper vanes. (b) Electric field strength along the center line of the cavity. (c) Distribution of the electric field lines in the transverse ($x$, $y$) plane at the kicker cavity center. (d) Simulated magnetic (left) and electric (right) field distributions in the middle ($x$, $y$) plane of the cavity. The maximum on-axis electric field strength is 11.5 MV/m. (e) Simulated magnetic (left) and electric (right) field distributions shown in a ($x$, $z$) cut of the cavity. The maximum electric field strength of 25.9 MV/m, found at one of the vane edges, is much less than the Kilpatrick criterion of 38.36 MV/m that we had set. The maximum magnetic field strength of 37.8 kA/m occurs at the thinnest part of the vane base.} \label{fig:kicker2}
 \end{figure}

The kicker cavity design is based on two parallel 158-mm-long, 36-mm-wide copper vanes separated by 20 mm as shown in Fig.\ \ref{fig:kicker2}. The vanes are straight and parallel along the longitudinal cavity axis, but slightly concave in the vertical direction (along $y$), as can be seen in Fig.\ \ref{fig:kicker2}(c) and (d). The kicker cavity is aligned with its axis $1^{\circ}$ tilted to the right with respect to the main accelerator axis. This way, electrons deflected by the first auxiliary magnet in front of the kicker cavity in two-color mode by $1^{\circ}$ enter the cavity along its longitudinal axis. A homogenous transverse horizontal RF electric field of up to 11.5 MV/m (Fig.\ \ref{fig:kicker}(b)), is formed between the vanes when RF power is supplied to the cavity. With wall losses calculated to be up to 36 W, a small cooling flow is sufficient to stabilize the temperature keeping the cavity on resonance. As there is no beam loading (coupling) between the electron bunches and the cavity modes, the kicker cavity can be conveniently powered up 10 to 20 $\mu$s before the arrival of the electron bunch train, resulting in a constant steady-state electric deflection field. As the RF phase is adjusted to be on crest, phase noise or drifts have negligible effect on the electron bunch deflection. RF power to the kicker cavity is supplied from a solid-state (LDMOS power transistors) pulsed amplifier (JEMA France, Haguenau, France) delivering up to 56 kW of in-pulse power. 

\subsection{FIR FEL electron transfer and beam dynamics}

The axis of the FIR FEL is rotated by $90^{\circ}$ with respect to the accelerator axis (see Fig.\ \ref{fig:overview}). Thus, electron bunches deflected by $4^{\circ}$ to the right at the kicker cavity section have to be steered to the left onto the FIR FEL axis by $94^{\circ}$. This is achieved by an isochronous achromatic bend formed by two $47^{\circ}$-deflection dipoles (DF's) framing two doublets of quadrupole magnets. The first and fourth as well as the second and third of these identical quadrupoles are electrically connected in series, respectively. Two additional quadrupole doublets have been installed in the FIR FEL electron beamline; the first one is located upstream of the first DF dipole, while the second one is placed on the FIR FEL axis between the second DF dipole and the FIR undulator. To free up the space needed for the installation of the latter quadrupole doublet, the midpoint of the optical cavity is shifted by 296 mm upstream with respect to the undulator center. 
As will be detailed below in Section \ref{sec:resonator}, the waist of the cavity mode coincides with the undulator center.

\begin{figure}
\includegraphics[width=0.48\textwidth]{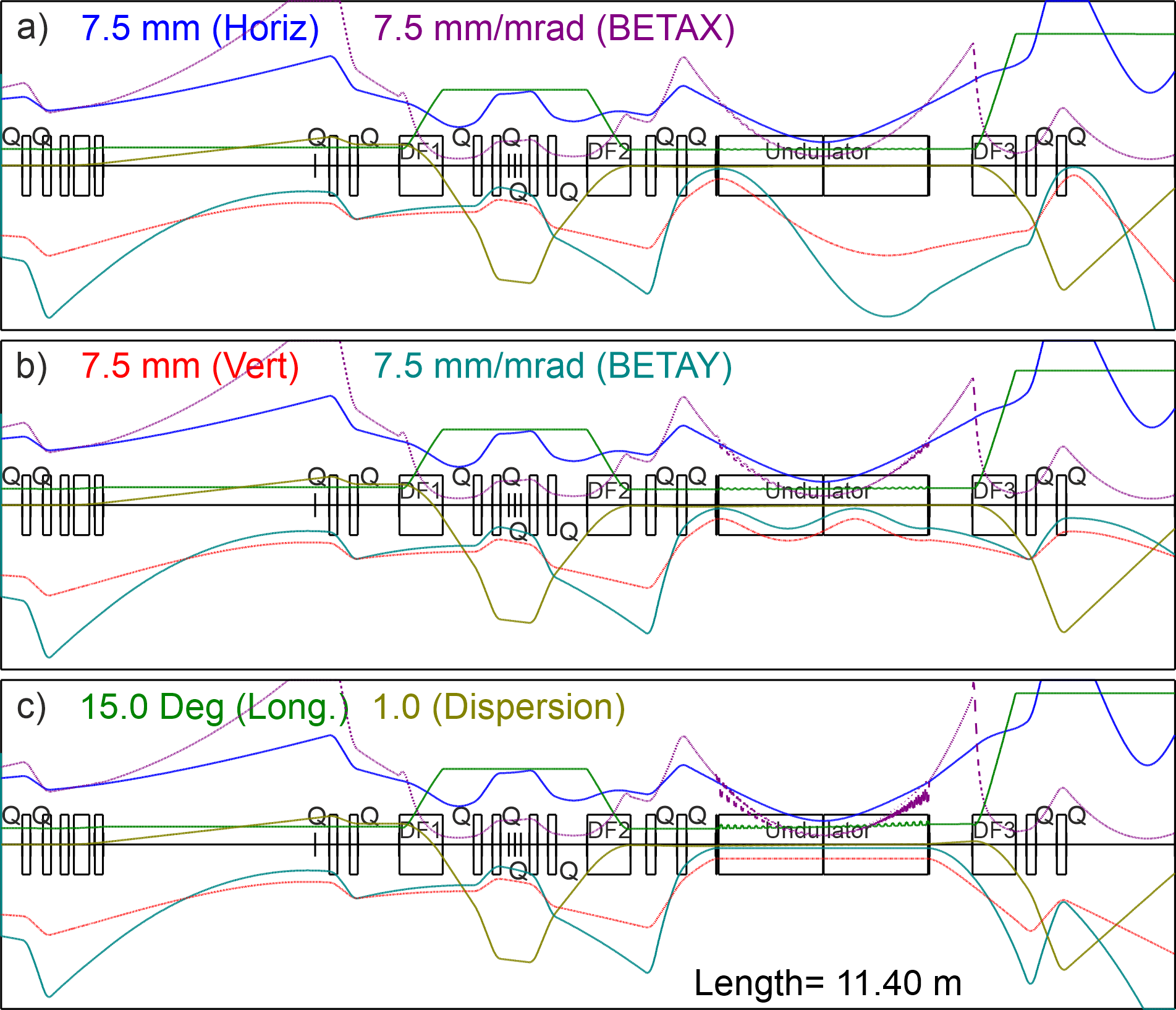}
\caption{TRACE-3D simulations of the FIR electron beam during an undulator gap scan at 18 MeV electron energy showing the loss of vertical focusing (red trace) as the FIR undulator gap opens from bottom to top: (c) minimum gap of 32 cm (corresponding to 162 $\mu$m wavelength at 18 MeV electron energy); (b) medium gap (53 $\mu$m); and (a) maximum gap (31 $\mu$m). The plots start at the left where the beam is focused by two quads (Q) downstream of the accelerator system; these are the quads QB05 and QB06 visible in Fig.\ \ref{fig:kicker}(b). The plots end to the right at the beam dump (not indicated). In between there are further quad doublets and the three $47^\circ$-deflection dipoles (DF1-3), as described in the text. The full scale of the vertical axis is indicated corresponding to the color code. Note that the horizontal and vertical dimensions are plotted in terms of $\sqrt{5}$ times the rms radius, i.e., the rms radii are 2.24 times smaller than the graphed values.} \label{fig:Trace_2}
 \end{figure}

The doublet ahead of the FIR achromat is used to capture the diverging beam, match it into the $94^{\circ}$ bend and ensure the FIR achromatic behavior, so it is not available for the FIR undulator input match. Instead, the doublet ahead of the undulator has to be used to match the beam into the undulator. Figure \ref{fig:Trace_2} shows TRACE-3D \cite{TRACE3D} plots of the FIR beamline performance for the most difficult low-electron-energy operation. Here the FIR electron beamline begins after the accelerator at the left edge of the plot and runs to the FIR FEL beam dump at the right edge. 
The plots, Fig.\ \ref{fig:Trace_2}(a)-(c), are for three different undulator gaps and otherwise identical parameters. As such, the traces all the way up to the undulator entrance are the same in the plots. The dispersion envelope (gold) goes to near zero after the FIR achromat indicating that the $94^{\circ}$ bend is indeed achromatic. 

The lowest of the three plots, Fig.\ \ref{fig:Trace_2}(c), has the undulator gap set to its minimum of 32 mm. The horizontal match into the undulator (blue) exhibits a waist in mid-undulator, while the near constant vertical envelope (red) illustrates the desired match into the vertically focusing undulator. The center plot, Fig.\ \ref{fig:Trace_2}(b), shows the situation with the undulator at the mid-point of a full gap scan, while the upper plot (a) is for the gap fully open. Given there is no undulator horizontal focusing, the horizontal beam remains well matched during a gap scan, but the vertical beam starts to exhibit betatron oscillations as the FIR undulator opens.  At maximum undulator gap (corresponding to 31 $\mu$m radiation at 18 MeV electron energy), the vertical match has deteriorated significantly, but the maximum vertical beam size is still less than 2 mm rms, which still fits nicely inside the larger FIR optical mode and lases adequately.  Note that the magnet settings ahead of the undulators are not changed for these simulations as the gap is opened. However, we have theoretically shown that scanning the pre-undulator quadrupole doublets in conjunction with the gap scan can keep the MIR and FIR FEL beams nicely matched into both undulators during simultaneous full gap scans. 

Note that the match into the FIR undulator also depends on the settings of yet another quadrupole doublet, namely the one after Linac 2 upstream of the kicker section (the first quadruple doublet at the left in Fig.\ \ref{fig:Trace_2}). However, in two-color operation, this quadrupole doublet is not freely adjustable for matching the electron bunches into the FIR undulator, because it also affects the matching into the MIR undulator. The settings of that quadrupole doublet used in Fig.\ \ref{fig:Trace_2}, in combination with the quadrupole doublet upfront of the MIR undulator and the MIR mid-achromat quadrupole triplet magnets, provide matching into the MIR undulator as well as enabling two-color operation.

\subsection{Short Rayleigh range oscillator} \label{sec:resonator}

Our design of the FIR FEL is based on a short-Rayleigh-range oscillator \cite{Colson_2006} which, in combination with the undulator described below, delivers FIR radiation from 4.5 to 175 $\mu$m. The Rayleigh range $Z_0 = 68$ cm equals about 1/3 of the undulator length which corresponds to a maximum in FEL radiation pulse energy, as discussed in Section \ref{sec:simulations}. A short-Rayleigh-range oscillator design implies a relatively large transverse optical-mode size at both ends of the undulator and even more so at the cavity mirrors \cite{Colson_2006}. The mode size scales with wavelength and, hence, becomes significant (many cm) in the FIR regime. A possible strategy to cope with an optical mode this large is installation of a one-dimensional waveguide for vertical confinement of the radiation. The main advantage of this approach is that it permits a reasonably small minimum undulator gap and correspondingly large undulator parameter as needed to get lasing at long wavelengths. However, waveguide effects have been reported to be detrimental to the performance of FIR/THz oscillator FELs because they lead to "spectral gaps", i.e.\ wavelengths where the output power is significantly reduced or lasing is suppressed all together. \cite{Prazeres2009,Ortega2014,Arslanov2014,Prazeres2016,Heting2022,Heting2023}.

\begin{figure}
 \includegraphics[width=0.48\textwidth]{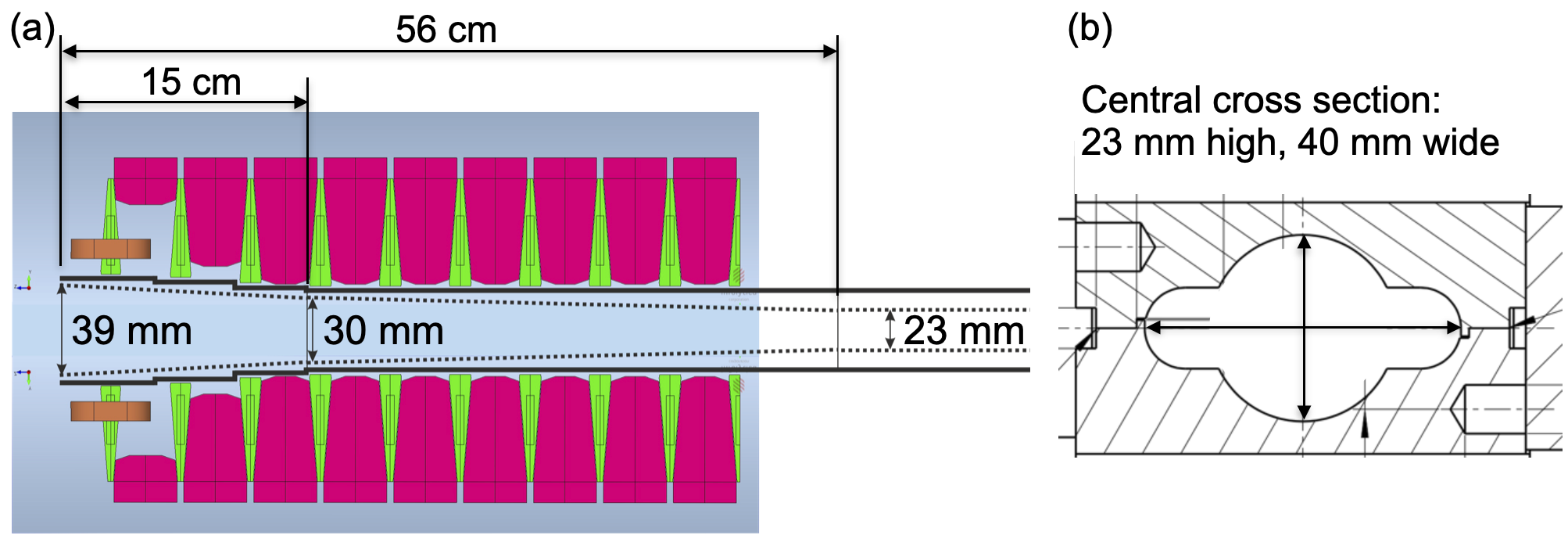}
 \includegraphics[width=0.48\textwidth]{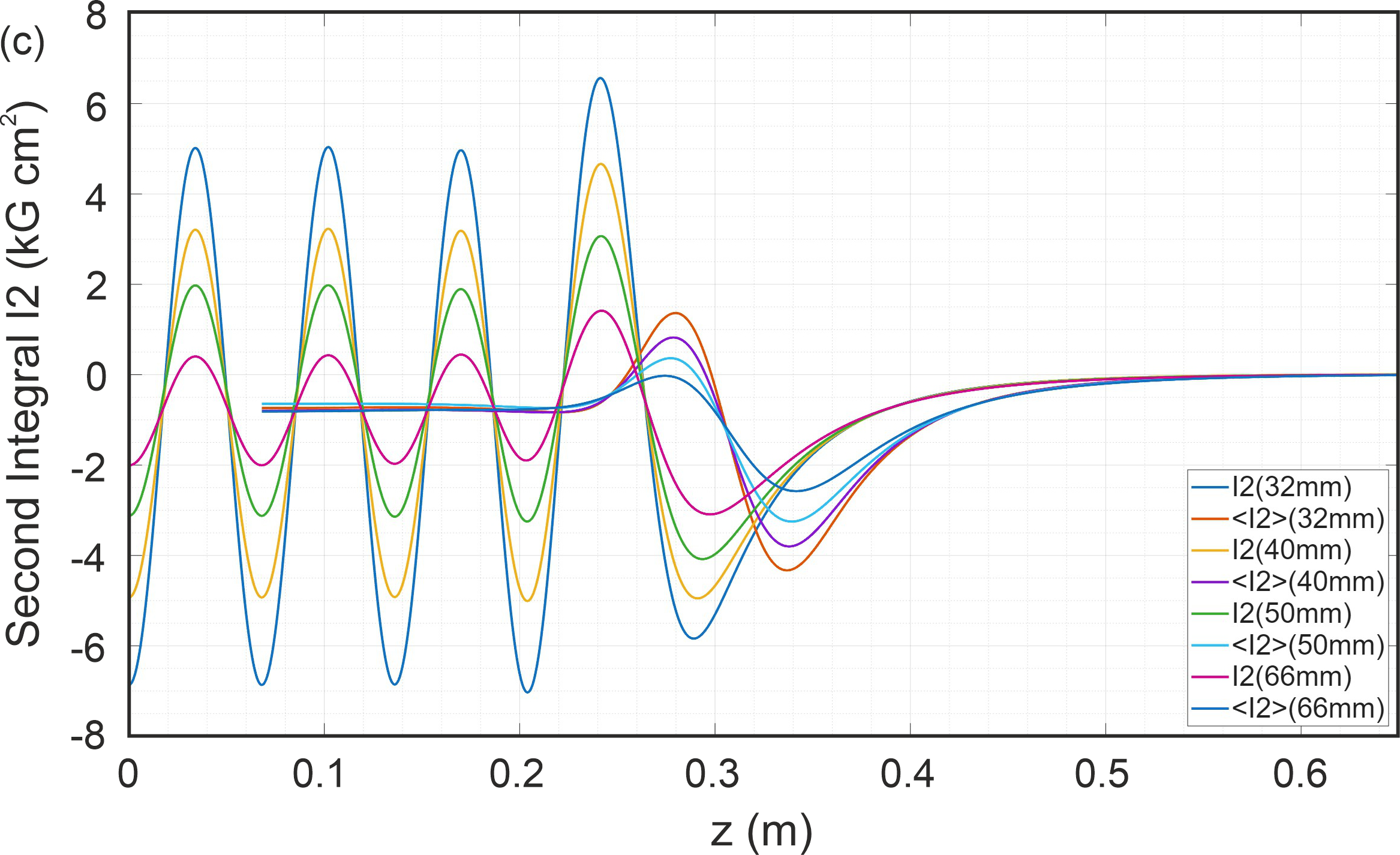}
\caption{(a) Cross sectional views of the FIR undulator vacuum chamber. The inner height of the chamber increases in two sections linearly first from 23 mm to 30 mm and then from 30 mm to 39 mm as indicated. The outer chamber height increases in three steps from 31 mm to 34 mm, then to 37.1 mm and eventually to 42.4 mm. Poles and magnets are recessed correspondingly at both ends of the undulator to accommodate for the vacuum chamber as can be seen in the longitudinal cross section. (b) Transverse cross section of the vacuum chamber at the undulator center. (c) Design trajectory, I2 (G\,cm$^2$), from finite-elements analysis for gaps of 32, 40, 50, and 66 mm with gap-dependent EM correction for the undulator entry. The average was taken over the undulator period of 68 mm.} \label{fig:U68chamber} 
 \end{figure}

To circumvent waveguide effects our design includes a large-clearance vacuum envelope throughout the 5.4 m long FEL resonator, thereby allowing for free-space propagation of the optical mode even at the longest wavelengths. To this end, the design of the 2.329-m-long undulator vacuum chamber, shown in Fig.\ \ref{fig:U68chamber}, shows an inner and outer height at the 1.2-m-long central section of as much as 23 and 31 mm, respectively. The inner height increases linearly towards both ends of the FIR undulator in two taper regions starting 153 and 565 mm from each end, as indicated in Fig.\ \ref{fig:U68chamber}(a). The horizontal inner width of the vacuum chamber is 40 mm throughout its full length. The vertically bi-tapered shape mimics the optical mode size variation along the undulator axis with its mode waist located at the undulator center. The vacuum pipes and chambers connecting the undulator chamber with the upstream and downstream cavity-mirror chambers have large inner diameters such that they do not clip the FIR optical mode. 

The nominal cavity length of the FIR FEL of $L_0 = 5.4$ m is identical to the MIR cavity length, which makes synchronization of MIR and FIR radiation pulses straightforward (see Section \ref{sec:crosscorrelation} below). The 8-cm-diameter concave spherical cavity mirrors are made out of copper with bare gold coating. Just as for the MIR cavity, we use hole-outcoupling for the FIR cavity. Each of four different downstream outcoupling-mirrors with hole diameters of 1.0, 2.5, 4.0, and 6.0 mm mounted on a precision in-vacuum translation stage can be positioned along the cavity axis and used as an outcoupling mirror. This makes it possible to chose the outcoupling-hole diameter that is best suited for a wavelength range of interest. The curvature radius of the outcoupling-mirror is 2596 mm, while the curvature radius of the upstream end mirror is 3150 mm. This defines the Rayleigh range of $Z_0 = 68$ cm and the shift of the resonator waist position by 296 mm with respect to the resonator center, thereby compensating for the undulator center being shifted by this amount towards the out-coupling end of the resonator. 
 
\subsection{FIR Undulator}
\label{sec:U68}

The IR wavelength range of the FIR FEL is defined by its undulator period $\lambda_{\rm U} = 68$ mm. This period was chosen such that at any electron energy set in the accelerator, the MIR FEL (at close to minimum undulator gap) and the FIR FEL (at close to maximum undulator gap) generate IR pulses at identical wavelength (3rd requirement described above). By opening up the MIR undulator gap and/or closing the FIR undulator gap the wavelengths of the FELs can be independently reduced and increased, respectively, over a wide tuning range (typically more than a factor of 2.5 in each FEL, see below). Furthermore, for the given range of electron energies accessible with the accelerator from nominally 15 to 50 MeV, the FIR FEL allows lasing from 4.5 to 175 $\mu$m, compliant with the users' request (1st requirement described above).

\begin{figure}
 \includegraphics[width=0.48\textwidth]{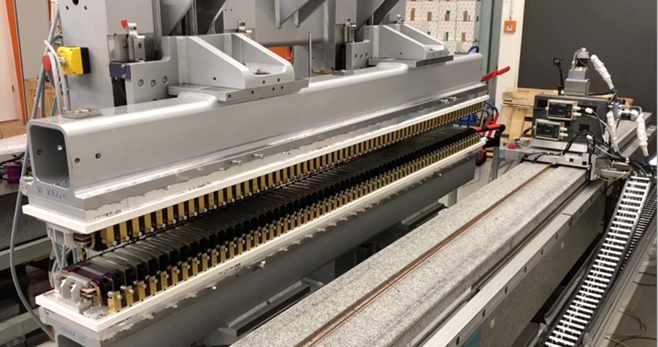}
\caption{Completed FIR undulator during magnetic mapping at FHI. The Hall probe scanner on its rigid 7-m-long granite slab is visible in the foreground.}\label{fig:U68Picture} 
 \end{figure}

As a consequence of the relatively large height of the FIR undulator chamber, the minimum gap of the FIR undulator needs to be as large as 32 mm. To achieve the high magnetic-flux density needed to get the large undulator parameter required for the long wavelengths, we implemented an optimized, radiation-resistant wedged-pole hybrid-magnet undulator design \cite{FEL_2012c}, schematically visible in Fig.\ \ref{fig:U68chamber}(a). The wedge-shaped permanent magnets were machined from a commercially available high-coercivity grade of NdFeB (VACODYM 983 DTP, Vacuumschmelze GmbH \& Co.\ KG, Hanau, Germany). This grade's manufacturing process includes dysprosium grain-boundary-diffusion (GBD) to optimize the magnetic properties. According to the manufacturer the improvements achieved by GBD are substantial; at the surface the intrinsic coercivity H$_{\rm cJ}$ as well as the 5\% and 1\% demagnetizing field strengths H$_{\rm d5}$ and H$_{\rm d1}$, respectively, increase by almost 50\% while remanence B$_{\rm r}$ and normal coercivity H$_{\rm cB}$ exhibit only 0.4\% reduction. The relevent numbers are summarized in Table\ \ref{tab:U68}. Considering the temperature dependence of H$_{\rm cJ}$, H$_{\rm d5}$ and H$_{\rm d1}$ this translates into a 50 - 60 K increase in safe operation temperature. This is a significant improvement in radiation hardness because electron showers can possibly raise domain temperature by about 150 K.

\begin{figure}
 \includegraphics[width=0.48\textwidth]{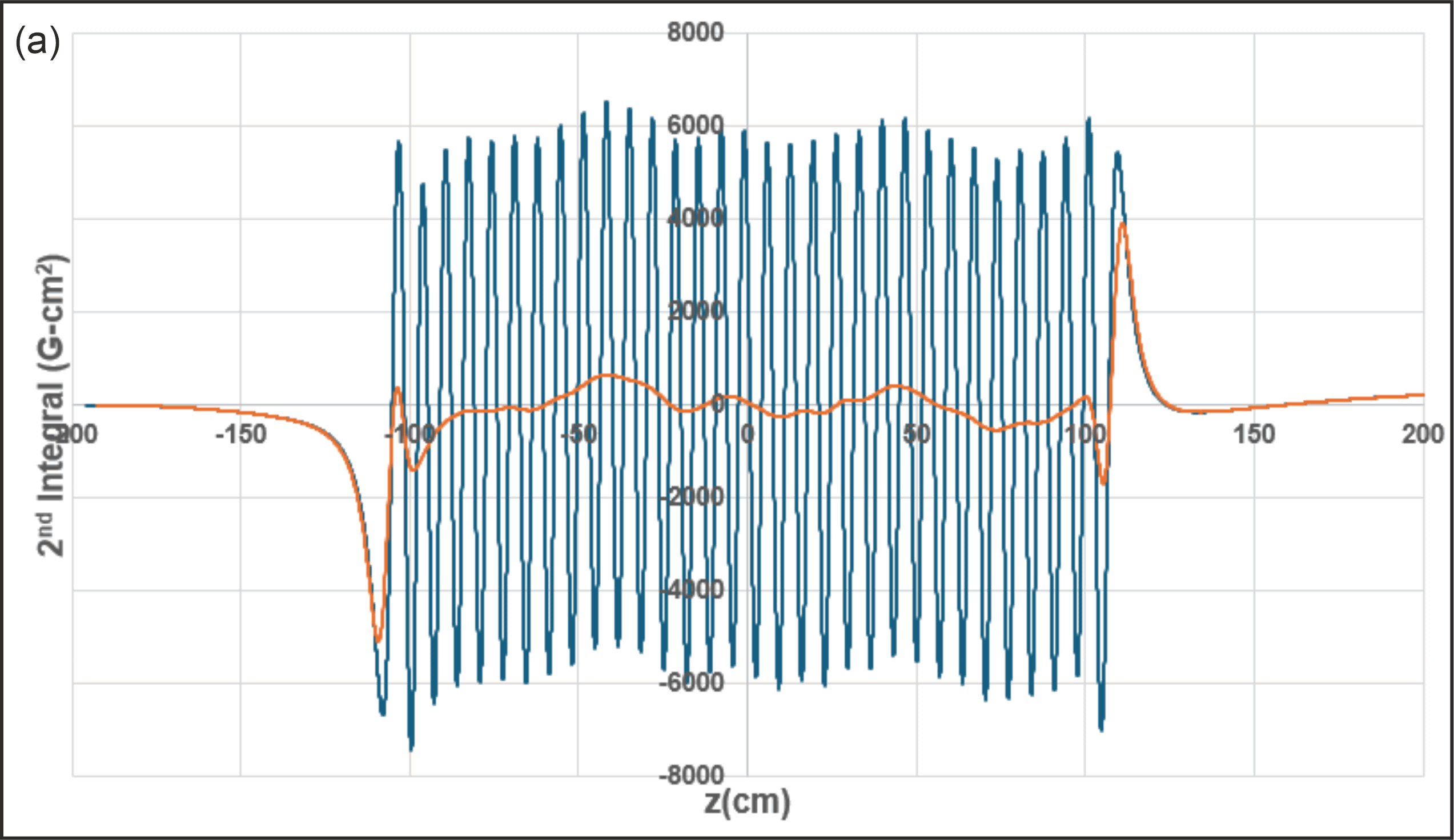}
 \includegraphics[width=0.48\textwidth]{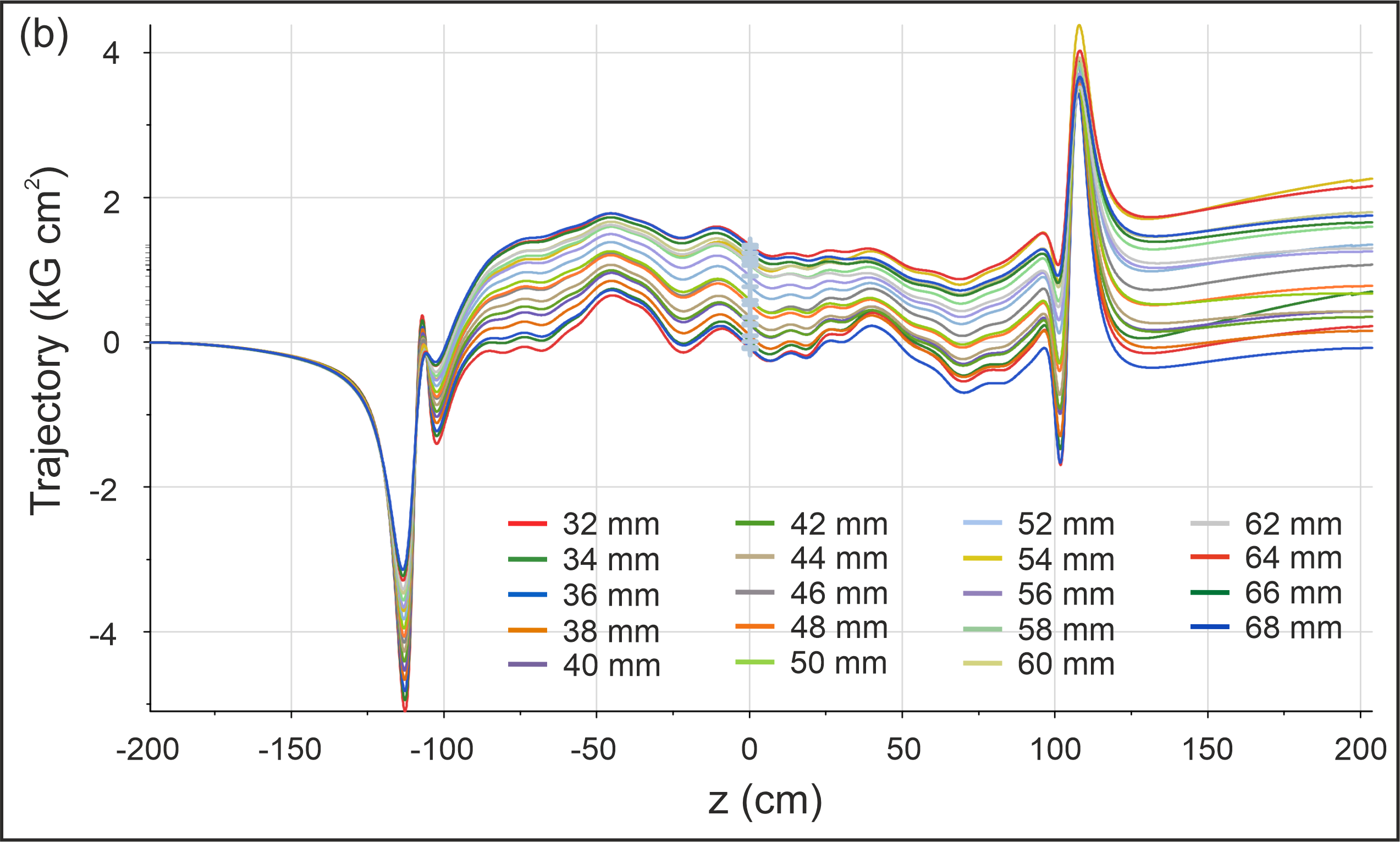}
 \includegraphics[width=0.48\textwidth]{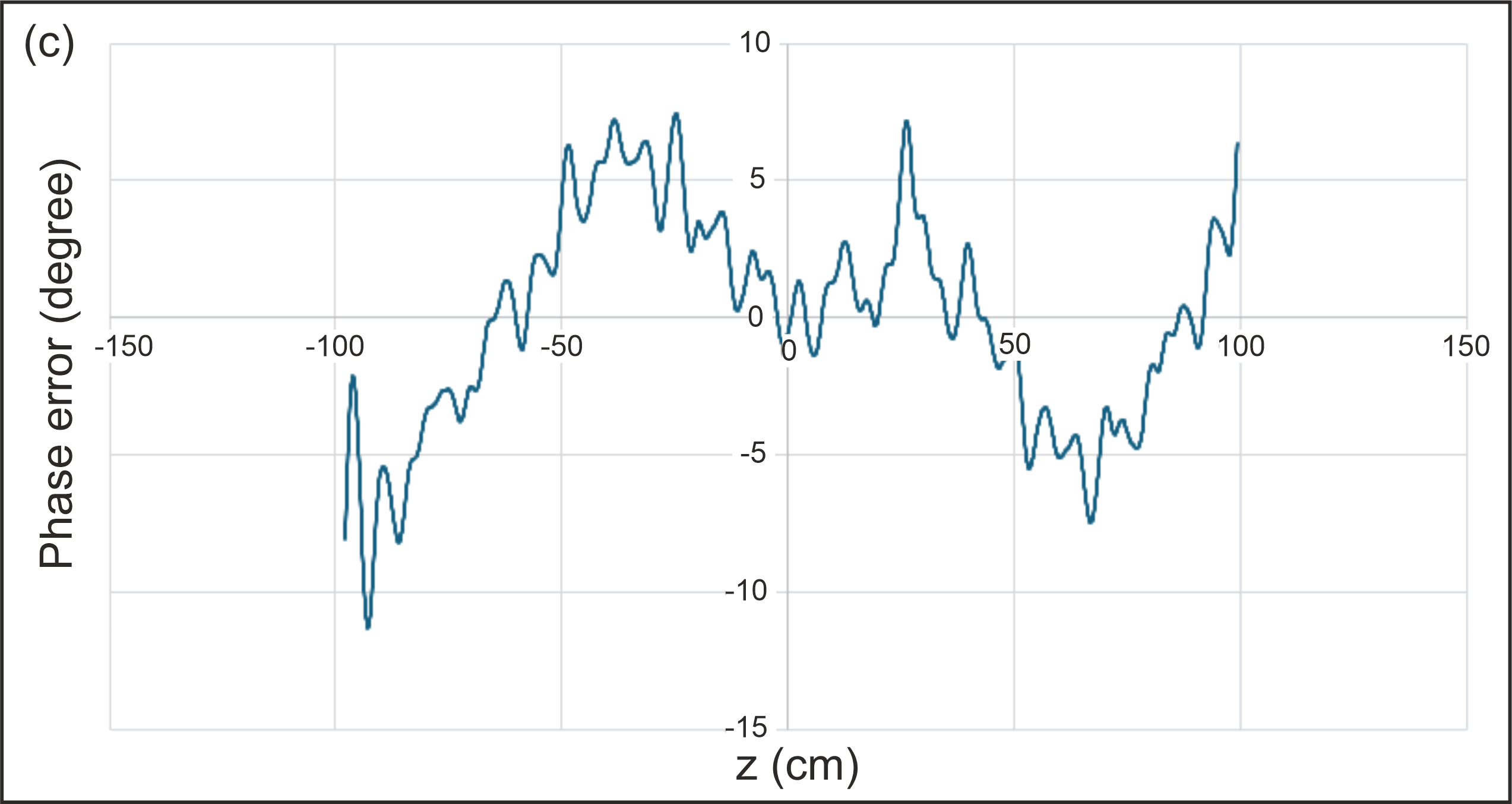}
 \includegraphics[width=0.48\textwidth]{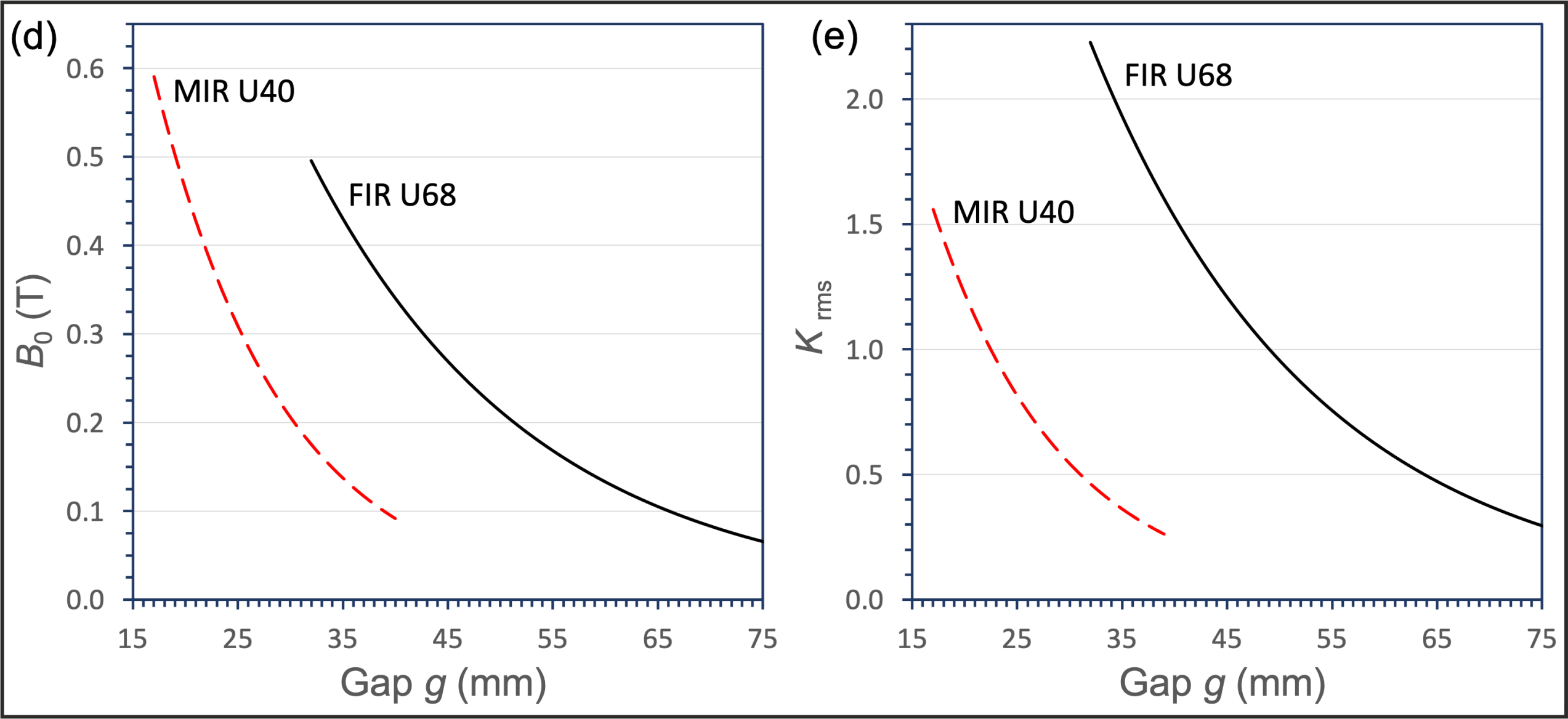}
\caption{Characterization of the FIR undulator. (a) 2nd integral of the FIR undulator derived from vertical-field measurements at the magnetic center line for minimum gap, $g = 32$ mm (blue line), and the same data averaged over the period length of 68 mm (red line). (b) FIR undulator vertical-field trajectories at the magnetic center line for various undulator gaps in 2 mm steps from 32 mm (red line, identical to the red line in (a)) to 68 mm (blue line). (c) Phase error. (d) Peak on-axis magnetic flux density $B_0$ and (e) root-mean-square undulator parameter $K_{\rm rms}$ versus gap for both undulators; MIR (dashed red lines) and FIR (solid black lines).} \label{fig:U68scans} 
 \end{figure}

The wedge-shaped poles were machined from the same grade of vanadium permendur that was previously used for the MIR undulator as well (VACOFLUX 50, Vacuumschmelze GmbH \& Co.\ KG, Hanau, Germany). It is characterized by a saturation polarization of 2.30 T. A comprehensive comparison of design and performance parameters of both the old MIR and new FIR undulators is given in Table\ \ref{tab:U68}.

%

\begin{table*}[hbt]
\small
  \caption{Design and performance parameters of the MIR and FIR hybrid-magnet wedged-pole undulators of the FHI FEL. First and second integral, phase error and roll-off were measured with a Hall probe scanner prior to installation in the vault.  (\textdagger) Magnetic properties of magnets and poles are provided based on measurements made by the manufacturer (Vacuumschmelze GmbH, Hanau, Germany). (\textdaggerdbl) Numbers in parenthesis refer to the 2nd and 1st magnet at each end of the undulator. ($^\mathsection$) Numbers in parenthesis refer to the 3rd (only FIR undulator), 2nd and 1st pole at each end of the undulator.}
  \label{tab:U68} 
 \begin{tabular*}{0.75\textwidth}{@{\extracolsep{\fill}}|r|r|r|} 
     \hline
                     & MIR undulator &  FIR undulator     \\
      \hline
      Year of commissioning  & 2012  &  2023      \\
      \hline
      Period $\lambda_{\rm U}$         & 40 mm           &  68 mm      \\
      \hline
      Pole and magnet shape       &   \multicolumn{2}{c|}{Wedged}     \\   
      \hline
      Wedge angle                       & 3$^\circ$         &  9$^\circ$             \\
      \hline
      Field symmetry       &   \multicolumn{2}{c|}{Anti-symmetric}     \\   
      \hline
      Total number of periods                    & 51   &  33    \\ 
      Number of optically useful periods   & 49   &  30    \\ 
     \hline
      Length (strongbacks)        &   \multicolumn{2}{c|}{2.286 m}     \\     
      \hline
      Gap range          &  16.5 - 150 mm    &   32.0 - 150 mm          \\       
       repeatability      &  $ < 5   \mu$m     &  $< 1   \mu$m    \\
       resolution          &  $ < 1   \mu$m     &  $< 1   \mu$m      \\
       bi-directional backlash & $< 1   \mu$m   &  $< 1   \mu$m    \\
     \hline
      Peak on-axis field at min.\ gap &  0.614 T  &    0.496 T            \\        
      \hline
      max.\ $K_{\rm rms}$ & 1.62     & 2.23              \\              
       \hline
      Magnetic material     &   \multicolumn{2}{c|}{Nd$_2$Fe$_{14}$B}     \\
      \hline
      Grade                    &  VACODYM 890TP  &  VACODYM 983DTP   \\
      Remanence$^\dag$ $B_{\rm r}$         &  1.18 T                      &   1.25 T                     \\
      Normal coercivity$^\dag$ $H_{\rm cB}$   & 919 kA/m            &     952 kA/m                                \\                    
      Intrinsic coercivity$^\dag$ $H_{\rm cJ}$   & 2625 kA/m          &     2230 kA/m (non-GBD zone)   \\                    
                                                        &                               &      2830 kA/m (GBD surface)          \\ 
      \hline
      Number of magnets                      & 2 x 101       &   2 x 65  \\
      \hline
      Magnet width                 &   85 mm           &  129.7 mm       \\
      height$^\ddag$              &  69.46 (65.35; 34.75) mm        &  61.3 (52.7; 22.7) mm     \\
      thickness at base           &  17.3 mm          &   30.7 mm        \\
      thickness at top$^\ddag $    &  13.9 (14.2; 15.8) mm          &   23.0 (24.4; 29.1) mm        \\
      \hline
      Pole material            &   \multicolumn{2}{c|}{Vanadium permendur}                                            \\
      Grade                       &   \multicolumn{2}{c|}{VACOFLUX 50 (49\% Fe, 49\% Co, 2\% V)}          \\
      Saturation flux density$^\dag$ $B_{\rm s}$   &   \multicolumn{2}{c|}{2.35 T (at $H = 40$ kA/m)}              \\
      Max.\ permeability$^\dag$ $\mu_{\rm r}$  &   \multicolumn{2}{c|}{6,700}                           \\
      \hline
      Number of poles      &     2 x 102       &   2 x 66  \\
      \hline
      Pole width                                       & 50.0 mm                            &  80.0 mm                                     \\
      height$^\mathsection$                    & 64.96 (64.46; 64.20) mm   &  52.1 (50.5; 48.7; 46.4) mm        \\
      thickness at base                            &  2.6 mm                             &  2.82 mm                                     \\
      thickness at top$^\mathsection$     &  5.9 (5.9; 5.9) mm             &  10.6 (10.5; 10.1; 9.8) mm          \\
      \hline
      First integral          & 13 G\,cm at $K_{\rm rms} = 1.5$    &  3 G\,cm at $K_{\rm rms} = 2.23$   \\
                                    & -37 G\,cm at $K_{\rm rms} = 0.5$   &  2 G\,cm at $K_{\rm rms} = 0.41$    \\
      \hline
      Second integral       & 1620 G\,cm$^2$ at $K_{\rm rms} = 1.5$    &     220 G\,cm$^2$ at $K_{\rm rms} = 2.23$     \\
                                      &       0 G\,cm$^2$ at $K_{\rm rms} = 0.5$    &   1750 G\,cm$^2$ at $K_{\rm rms} = 0.41$  \\
      \hline
      Phase error           &    $4^\circ$  at $K_{\rm rms} = 1.5$ &   $3.8^\circ$  at $K_{\rm rms} = 2.23$      \\
                                    & $1.9^\circ$  at $K_{\rm rms} = 0.5$ &   $1^\circ$  at $K_{\rm rms} = 0.41$        \\
      \hline
      Transverse roll-off  & 0.7\%  at $K_{\rm rms} = 1.5$    &  0.2\% at $K_{\rm rms} = 2.23$      \\      
      at $x=10$ mm        &  1.6\%  at $K_{\rm rms} = 0.5$   &  0.8\% at $K_{\rm rms} = 0.41$     \\
      \hline
      Peak-to-peak trajectory  &  $< 2000$  G\,cm$^2$       &  $< 1680$ G\,cm$^2$     \\
      deviation at any gap      &                         &     \\
      \hline
      rms trajectory deviation at any gap &        $< 550$ G\,cm$^2$       &      $< 300$ G\,cm$^2$    \\
      \hline
\end{tabular*}
\end{table*}

The bi-tapered undulator vacuum chamber shows a stepwise outer-height increase at both ends as shown in Fig. \ref{fig:U68chamber}(a). To accommodate the chamber, the tops of the first three poles and the first two magnets of the FIR undulator are recessed at both ends of the undulator. This was achieved by installing reduced-height poles and magnets as indicated in the drawing shown Fig. \ref{fig:U68chamber}(a). As a result of this design, the first 1\textonehalf \ periods with tapered magnetic fields at each end of the undulator guide the electrons in and out, but they are not involved in the FEL interaction. As such, the number of periods active in the FEL interaction is reduced to $N=30$ from a total of 33 periods. 

A relatively short undulator is necessary to fit within the short Rayleigh range oscillator mode described in Section \ref{sec:resonator} in order to avoid clipping on the inside of the beam pipe at the ends of the undulator at large wavelengths. The resulting small number of undulator periods has two additional advantages: greater energy extraction proportional to $1/N$, and reduced slippage of the radiation pulse over the electron pulse also proportional to $1/N$. One disadvantage is reduced weak field gain proportional to $N^3$, but as the gain exceeds output coupling loss, the FEL will start and operate at strong field saturation.

Assembly of the undulator was done at FHI where also magnetic field measurements were performed using a 7 m monolithic-granite Hall-probe (Model YM12, Senis AG, Baar, Switzerland) magnetic-field scanner (visible in Fig.\ \ref{fig:U68Picture}). Magnetic center line vertical field trajectories versus undulator gap are presented in Fig.\ \ref{fig:U68scans}. The spikes are at the entry and exit of the undulator. The main effect is a steadily increasing offset with gap caused by end fields. 

The peak on-axis magnetic-flux density $B_0$ = 0.49 T at minimum gap corresponds to a maximum root-mean-square undulator parameter $K_{\rm rms}$ = 2.23. Measurements of the peak on-axis magnet field of the undulator as a function of undulator gap are well fitted by an exponential expression;
\begin{equation}
B_0  = a \exp{\left( -b \frac{g}{\lambda_{\rm U}}\right) } \ ,
\end{equation}
where $g$ denotes the undulator gap. For the FIR undulator the best fit is found for $a = 2.224$ T and $b = 3.19$. The best-fit curves for $B_{0}$ and the root-mean-square undulator parameter $K_{\rm rms}$, which is given by $K_{\rm rms} = (93.36/\sqrt{2}) B_0 \lambda_{\rm U}$ ($B_0$ in Tesla and $\lambda_{\rm U}$ in meter) are plotted in Fig.\ \ref{fig:U68scans}(d) and (e), respectively. The corresponding curves for the MIR undulator (with $a = 2.34$ T and $b = 3.24$) are plotted for comparison. 

%
%

\section{FIR FEL Performance}
 \label{sec:FIR}
 
\begin{figure}
 \includegraphics[width=0.48\textwidth]{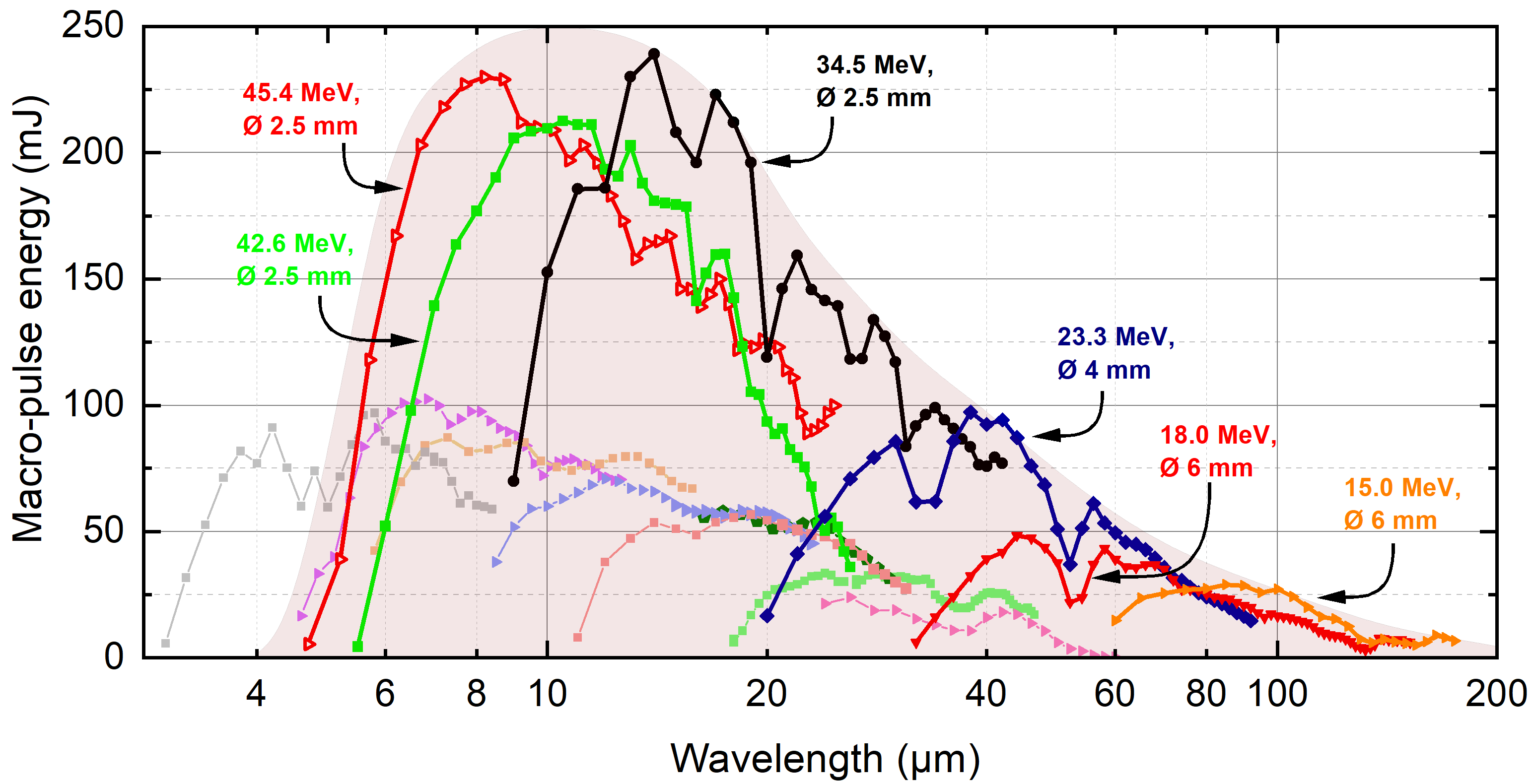}
\caption{Undulator gap-scan curves of the FIR and MIR FELs displaying the macro-pulse power plotted on a logarithmic wavelength axis. The upper bright-color traces indicate the pulse energy measurements of the FIR FEL at the indicated electron energies, while the lower faded-color curves, replotted from Fig.\ \ref{fig:MIRpwr}, show the results of the MIR FEL for comparison. The measurements were done at 1 GHz micro-pulse repetition rate and for narrow-spectral conditions. Different outcoupling-hole diameters of 2.5, 4, and 6 mm were used as indicated. The shaded area serves as a guide to eye and indicates the achievable macro-pulse power for the range of electron energies.} \label{fig:PWRcurves}
 \end{figure}

First light from the FIR FEL beamline was achieved at a wavelength of 8 $\mu$m in June 2023. To characterize the FIR FEL, we subsequently measured power curves at different electron energies between 15 and 45 MeV. The plots in Fig.\ \ref{fig:PWRcurves} show macro-pulse energies plotted as a function of wavelength on a logarithmic abscissa. In each curve, the wavelength was tuned by changing the undulator gap at the indicated electron energy. The MIR FEL measurements presented above in Fig.\ \ref{fig:MIRpwr} are replotted here for comparison. Just as with the MIR data, the FIR measurements were done for a relatively narrow spectral width of $\Delta \lambda / \lambda \approx 0.5(\pm 0.2)$\% achieved by appropriate adjustment of the cavity length detuning $\Delta L$. Higher pulse energies can be achieved for smaller $\Delta L$ and correspondingly larger spectral width.

The six curves of the FIR FEL plotted in Fig.\ \ref{fig:PWRcurves} indicate continuous lasing from 4.7 $\mu$m to 175 $\mu$m. For each curve the tuning range is at least a factor of 3, almost a factor of 5 for some curves. This large continuous tunability of the FIR FEL is particularly advantageous for many user experiments. In addition, the maximum pulse energy of the curves ranging from 50 mJ at low electron energy to more than 200 mJ at 45 MeV appears to be more than the corresponding values observed with the MIR FEL. The short Rayleigh range design of the FIR FEL with a smaller number of active periods, $N=30$, extracts more energy from the electron pulses, thus more energy to the optical pulses, compared to the MIR FEL with its larger number of undulator periods, as discussed in Section \ref{sec:U68}. This observation is in agreement with theory and simulations. 

Finally, it is noteworthy to mention that dips in the power curves are visible at wavelengths of 23, 33, 51, 93 and 133 $\mu$m. In particular, the ones appearing at 33 and 51 $\mu$m are significant. The origin of the dips is not yet understood. More experiments and/or simulations will be needed to understand the cause of the dips and, if possible, to get rid of them.
 
 \subsection{Simulation of FIR FEL Performance}
 \label{sec:simulations}
    
\begin{figure}
 \includegraphics[width=0.47\textwidth]{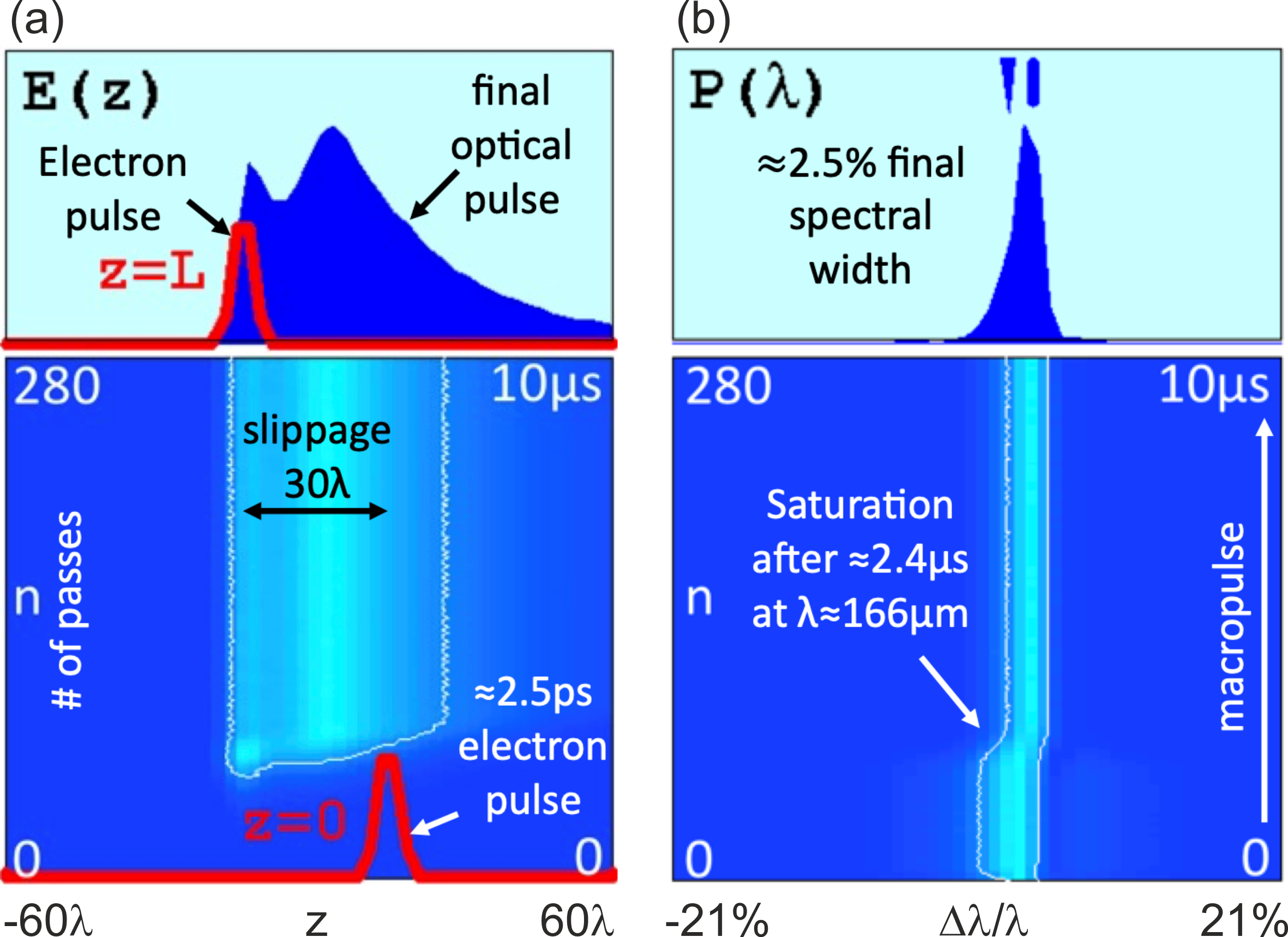}
\caption{Simulated FIR output radiation performance at a long wavelength of 166 $\mu$m, an electron energy of 18 MeV, 200 pC bunch charge and 2.5 ps bunch length. (a) Optical pulse shape: (lower part) evolution of the optical pulse over 280 round trips (corresponding to 10.08 $\mu$s) in the FEL cavity and; (upper part) final micro-pulse shape at the end of the macro pulse. The red trace indicates the electron bunch position relative to the optical pulse at the entrance (z = 0) and exit (z = L) of the undulator. (b) Optical spectrum: (lower part) evolution of the spectrum over 280 round trips in the FEL cavity and; (upper part) final spectrum. 
}  \label{fig:Bill}
 \end{figure}

Both the FHI MIR and FIR FEL oscillators are theoretically analyzed with 4D simulations using self-consistently coupled Maxwell-Lorentz equations including electron pulse interaction with the co-propagating light pulse. FELs operating in the FIR often suffer serious gain reduction due to short-pulse slippage. This occurs when a short electron pulse, in the range of 5 ps "length" (corresponding to an actual length $L_B = 1.5$ mm), is completely passed over by the FIR radiation pulse during one pass through the undulator. At FEL resonance, by definition, one wavelength of light $\lambda$ passes over an electron as the electron travels through one undulator period. For an undulator of $N$ periods, the slippage distance of light over the electrons is $N \lambda$. The severity of slippage is quantified by the dimensionless quantity $\sigma_z = L_B / N \lambda \approx 1.5  {\rm mm}/N \lambda$, or simply $\sigma_z  \approx 10/N$ for $\lambda =150$ $\mu$m. For $\sigma_z < 1$, much of the optical pulse passes over the electron pulse each pass and is not amplified. Weak-field gain and strong field extraction are both significantly reduced when $\sigma_z \ll 1$. 

Surprisingly, the FEL can work with short pulses, $\sigma_z < 1$, but the detailed interaction is complicated and is evaluated numerically in 4D simulations in (x,y,z,t) \cite{Blau:2015}. The longitudinal dimension in the 4D simulations is used to evaluate the reduction in performance due to slippage as the electron and optical pulses slip over each other. In the transverse direction, the radiation mode can also be complex due to the hole out-coupling and interaction with an electron beam with a radius much smaller than the optical mode. The numerical methods used follow the optical and electron pulses in (x,y,z,t) with no restrictions on the number of longitudinal or transverse modes other than the number of sites used. The methods have been compared to many experimental observations over four decades, and more importantly are consistent with analytic results \cite{Blau:2015}.

The Rayleigh length of the resonator $Z_0$ is defined by the radius of curvature of the gold mirror surfaces, as described in Section \ref{sec:resonator} above, and influences the FEL interaction. Not only does the Rayleigh length determine the mode waist, it also determines how rapidly the wavefronts expand by diffraction away from the mode waist $w_0$ at $Z = 0$. The fundamental mode radius is given by $w = w_0 (1 + Z^2/Z_0^2)^{1/2}$, where $Z_0 \lambda = \pi w_0^2$, and the radius ($w$, $w_0$) is defined as the radius where the intensity drops to 1/$e^2$. For $Z_0 \simeq L_{\rm U}/\sqrt{12} \approx 0.3 L_{\rm U}$, the optical mode volume along the undulator of length $L_{\rm U}$ is minimized so that light couples well to the enclosed electron beam. The mode in the 4D simulations is free to evolve and does not exactly stay in the fundamental mode. The true FEL optical mode finds a steady-state shape determined by a combination of the electron beam interaction, resonator mirrors, and beam pipe clipping. If the Rayleigh length is too large, the optical mode waist tends to be large and leads to excess clipping and diminished FEL interaction. If the Rayleigh length is too small, the wavefronts expand rapidly and clipping is increased at each end of the undulator. Clearly, this suggests that there is an optimum value of $Z_0$ that maximizes the FIR pulse energy. 

Figure \ref{fig:Bill} shows results of numerical 4D simulations for 18 MeV electron energy and $K_{\rm rms}  = 2.29$ at a FIR wavelength of 166 $\mu$m. Although slippage is large at these conditions, $N \lambda = 4.98$ mm, the numerical results indicate sufficient gain resulting in saturation reached after 2.4 $\mu$s. The optical pulse E(z) is $\approx 44\, \lambda$ long, giving a moderately wide spectrum, $\Delta \lambda / \lambda \simeq 2.5$\%, at moderate cavity detuning of $3 \lambda$. Larger cavity detuning would give a narrower, stable spectrum if desired.

\section{Two-Color Performance}
\label{sec:2color:performance}

\begin{figure}
 \includegraphics[width=0.47\textwidth]{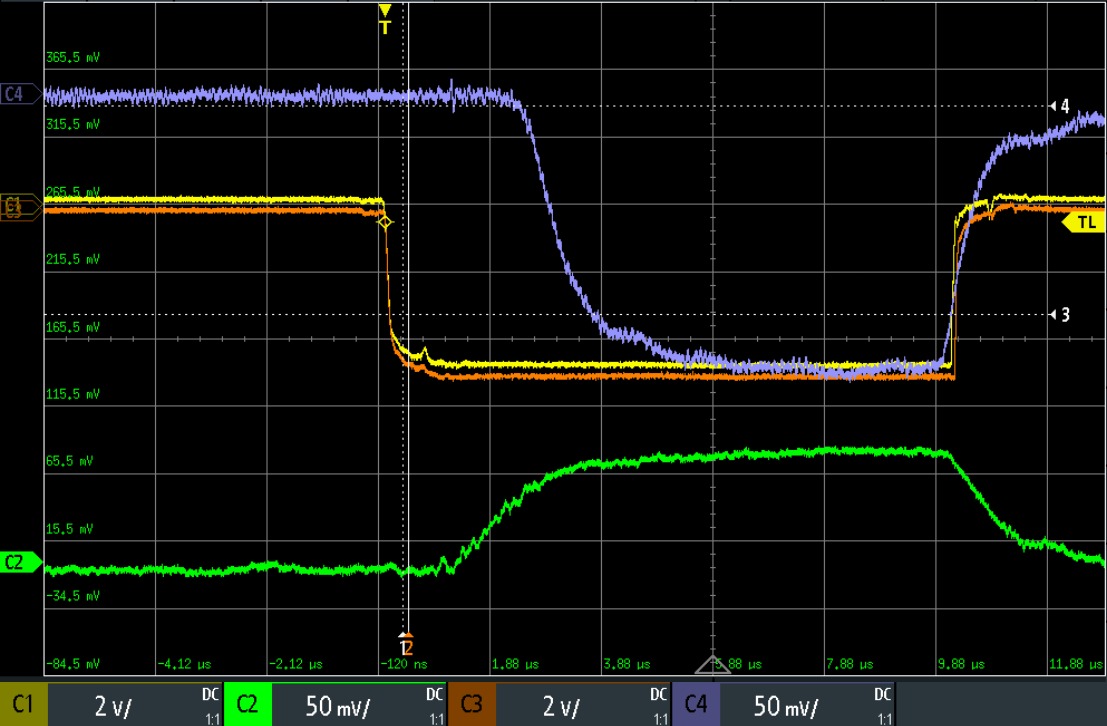}
\caption{Oscilloscope traces showing the first simultaneous two-color lasing from the MIR and FIR beamlines, each one operating at 500 MHz micro-pulse repetition rate. The yellow and orange traces show the 10-$\mu$s-long macro-bunches in the dumps (micro-bunches are not resolved). The green and violet traces show the optical signals of the MIR and FIR FEL, respectively, recorded with pyro detectors. Here, too, the micro-pulses are not resolved. The opposite sign of the radiation traces is a technical artefact. The horizontal scale is set to 2 $\mu$s per division.} \label{fig:osci} 
 \end{figure}

\begin{figure}
 \includegraphics[width=0.48\textwidth]{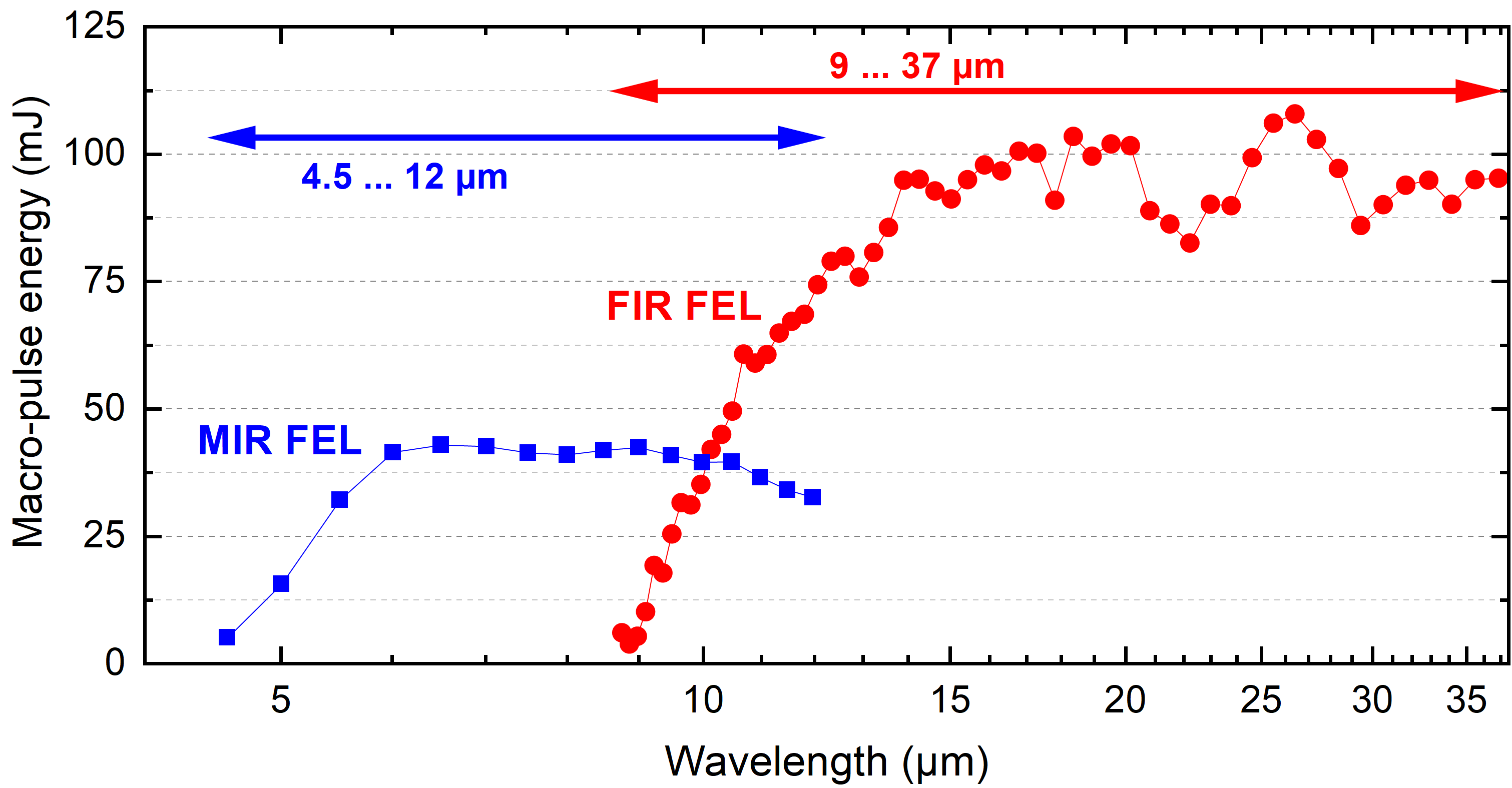}
\caption{First two-color, separately tunable, simultaneous lasing gap scans observed at an electron energy of 36.6 MeV. The plot indicates the overlap of the wavelength ranges at 9 to 12 $\mu$m (fulfilling the 3rd user  requirement, see text) and the wide, continuous tuning capabilities of both FELs.} \label{fig:2colorpwr} 
 \end{figure}

First simultaneous lasing of the MIR and FIR FELs was observed for an electron energy of 22.6 MeV at wavelengths of 18 and 55 $\mu$m, respectively, 
thereby achieving the project goal of two-color lasing. Figure  \ref{fig:osci} shows the lasing traces from that milestone. Subsequently, the first separately tunable, two-color lasing gap scans were obtained at 36.6 MeV electron beam energy, as shown in Fig.\ \ref{fig:2colorpwr}. A factor of four in the wave-length tuning range can be seen for the FIR radiation spectrum (9 to 37 $\mu$m) and nearly a factor of three for the MIR spectrum (4.5 to 12 $\mu$m). This corresponds to a tuning range of almost a factor of 10 for the ratio of FIR to MIR wavelength; it can be continuously varied from 0.75 to 7.0. The FIR pulse energy is a substantial 100 mJ, which would deliver 200 $\mu$J in 1 GHz FIR-only lasing mode.

As described above, the standard mode of operation of the accelerator system is a 1 GHz bunch repetition rate, which, in the two-color mode, results in a repetition rate of 500 MHz in each FEL. The data plotted in Fig.\ \ref{fig:2colorpwr} was measured in this mode. Reduced repetition rates in both FELs are possible in two-color operation as well. For instance, a bunch repetition rate of 55.6 MHz in both MIR and FIR pulses, as needed for some applications, has been implemented. In this mode, the bunch repetition rate of electron gun accelerator is set to 111.1 MHz corresponding to 9 ns bunch-to-bunch separation. Note that 9 ns (and, in general, any odd number of nanoseconds) leads to a left-right bunch separation in the kicker cavity, because the latter operates at 500 MHz, i.e., 2 ns period. For an even number of nanoseconds, however, the kicker cavity would deflect all the bunches arriving from the accelerator either to the MIR or to the FIR FEL. As such, from the possible reduced repetition rate modes of accelerator-system operation listed in Table\ \ref{tab:rates}, only those two leading to odd bunch-to-bunch separations (namely 3 ns and 9 ns), are viable modes of two-color operation. 

\subsection{Cross Correlation Measurements}
\label{sec:crosscorrelation}

The MIR and FIR pulses are generated independently in two oscillator FELs by electron bunches selected alternately from the bunch train produced by the accelerator system. Considering the high-precision timing of the accelerator and the low pulse-to-pulse jitter observed previously for the MIR pulses \cite{Kiessling_2018} we expect the MIR and FIR pulses to be synchronized on the sub-ps time scale. To check this we use a cross-correlation setup with the MIR and FIR pulses overlapping in a GaSe nonlinear crystal where the material's second-order response leads to sum frequency generation (SFG).

\begin{figure}
\includegraphics[width=0.48\textwidth]{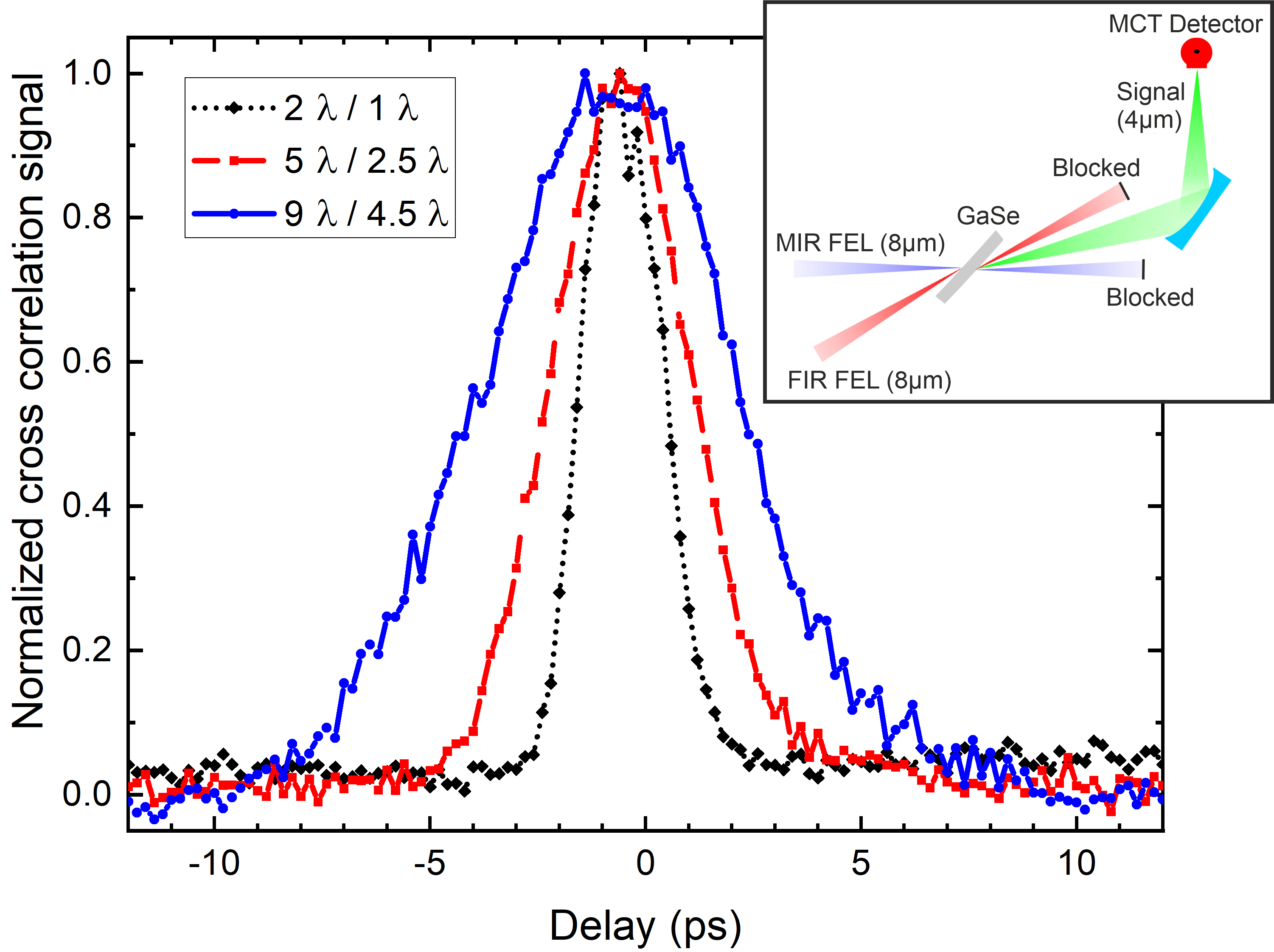}
\caption{Cross-correlation measurements of MIR and FIR FEL micro pulses. The sum-frequency signal generated by the nonlinear response of a GaSe crystal is measured as a function of FIR-to-MIR delay for three different combinations of FEL cavity length detunings: (black dotted) $2 \lambda$ MIR, $1 \lambda$ FIR; (red, dashed) $5 \lambda$ MIR, $2.5 \lambda$ FIR, and (blue, solid) $9 \lambda$ MIR, $4.5 \lambda$ FIR. The inset shows a schematic of the cross-correlation setup. All data were acquired with 8 $\mu$m central wavelength of both FELs with the accelerator running at reduced electron-bunch repetition rate of 111.1 MHz corresponding to 55.6 MHz pulse repetition rates in each MIR and FIR (see Table \ref{tab:rates}).} \label{fig:crosscorrelation}
\end{figure}

The SFG signal is measured as a function of delay between the MIR and FIR pulses (Fig.\ \ref{fig:crosscorrelation}). It reaches a maximum for zero delay corresponding to identical arrival times of the MIR and FIR pulses at the GaSe crystal. The optical path lengths for the MIR and FIR pulses propagating in separate evacuated IR beamlines from the corresponding FEL cavity out-coupling mirrors in the vault all the way to the cross-correlation end station are different; 54.04 m for the MIR and 52.61 m for the FIR. In addition, the electron beamlines from the accelerator to the MIR and FIR FEL cavities have different path lengths, as shown in Fig.\ \ref{fig:overview}. Both effects result in an arrival time difference of the MIR and FIR pulses at the optical setup on the order of 10 ns. While this is negligible in regard of temporal overlap of MIR and FIR macro pulses it needs to be compensated for overlapping micro pulses. To this end, the path lengths on the optical table were adjusted such that fine-tuning the temporal overlap of MIR and FIR micro pulses at the GaSe crystal can be achieved by using a precision translation stage in the FIR path that allows for precise adjustment of the MIR-to-FIR delay by $\pm 2$ ns.

The 1 mm thick \emph{z}-cut GaSe crystal is placed in the position of the overlapping MIR and FIR beam focii with incidence angles of 27$^\circ$ and 57$^\circ$, respectively, with respect to the surface normal, as shown schematically in the inset of Fig.\ \ref{fig:crosscorrelation}. This setup corresponds to the type II \emph{eoe} phase-matching condition \cite{Kiessling_2018}. The MIR and FIR pulses transmitted by the crystal are blocked geometrically while the SFG pulses, emitted at an angle in between, are detected by a mercury-cadmium-telluride (MCT) detector (InfraRed Associates, Inc., Stuart, FL, USA), similar to previous work of non-collinear FEL second harmonic generation spectroscopy \cite{Paarmann_2015}.

Fig.\ \ref{fig:crosscorrelation} shows cross-correlation signals as a function of the MIR-FIR pulse delay measured for three different combinations of FEL cavity length detuning $\Delta L$. In each case the cross-correlation peak width (full width at half maximum) is on the order of picoseconds ranging from 2.1 ps for small detuning of $\Delta L = 2 \lambda$ (MIR), $1 \lambda$ (FIR) via 3.7 ps for intermediate detuning of $5 \lambda$ (MIR), $2.5 \lambda$ (FIR) to 7.1 ps for $9 \lambda$ (MIR), $4.5 \lambda$ (FIR). Notably, for the given values of detuning, we expect MIR pulse durations in the range of 1 to 5 ps \cite{Kiessling_2018}. Due to the reduced number of undulator periods in the FIR undulator, similar pulse durations are expected for approximately half the detuning for the FIR pulse compared to the respective MIR value \cite{Colson_1990}, as confirmed by the nearly symmetric curves in Fig.\ \ref{fig:crosscorrelation}. Thus, cross-correlation widths on the order of the expected individual pulse durations suggest a maximum timing jitter between MIR and FIR micro pulses in the sub-ps regime. This confirms the high degree of synchronization inherently present in the two-color dual cavity FEL operated with a single accelerator system. A more detailed investigation with additional measurements of MIR and FIR FEL pulses cross-correlated with short pulses from a table top laser synchronized to the FEL is underway.

\section{Conclusions}
\label{sec:conclusions}

To date, the newly installed FHI FIR FEL has delivered radiation from 4.7 to 175 $\mu$m as shown in Fig.\ \ref{fig:PWRcurves}, using electron beam energies ranging from 17 to 45 MeV. We expect to further extend this wavelength range by tuning the accelerator to its full nominal energy range from 15 to 50 MeV. The pulse energies of the FIR are larger than those we obtained with the MIR FEL. We attribute this behavior, which was predicted by simulations, to the short-Rayleigh-range design of the FIR FEL.

In addition, we have demonstrated operation of 2-color lasing by running both MIR and FIR FELs simultaneously. In this two-color mode, it is possible to tune the FIR-to-MIR wavelength ratio from 0.75 to 7.0, i.e.\ almost by a factor of 10 via undulator-gap scan in each FEL. The 2-color mode is made possible by a 500 MHz side-deflecting cavity steering every second electron bunch to the MIR and every other second to the FIR FEL. In this way, each of the two FELs generates radiation pulses at a repetition rate of 500 MHz. The two-color mode works at reduced repetition rates as well; for instance, when operating the accelerator at an electron bunch repetition rate of 111.1 MHz, MIR and FIR radiation pulses are generated at 55.6 MHz repetition rate each. The MIR and FIR pulses are highly synchronized, because they are generated by electron bunches picked alternately out of a single bunch train from the accelerator. This has been confirmed by cross-correlation measurements indicating that the remaining MIR-FIR pulse-to-pulse timing jitter is in the sub-ps regime. 

First two-color operation of an IR FEL was demonstrated in pioneering work by J.-M.\ Ortega and coworkers at the CLIO facility in Paris Orsay in the 1990's \cite{JAROSZYNSKI:1994aa,JAROSZYNSKI:1995ab,Prazeres:1998ac}. In those experiments two identical undulator segments within the CLIO cavity were independently gap tuned, referred to as step-tapered undulator. Based on the strong gain of CLIO it was possible to observe two-color operation with a single cavity IR FEL. However, using this concept, lasing in the first undulator segment interferes with lasing in the second one leading to power fluctuations during gap tuning. In addition, the wavelength ratio was limited to a factor of ~1.5. The two-color mode of a dual-cavity IR FEL presented in this work overcomes those limitations and makes it possible to provide users with the wide and continuous tunability of both wavelengths that is needed for many applications. 

\begin{acknowledgments}
We are grateful to Ulf Lehnert and Lex van der Meer for highly fruitful discussions. We thank Henrik Loos for his contribution to the machine control software and Arjan van Vliet for his support during undulator assembly. A.C.B. acknowledges funding from the Max Planck-Radboud University Center for Infrared Free-Electron Laser Spectroscopy. A.Y.T.B. acknowledges support from the IMPRS for Elementary Processes in Physical Chemistry.
\end{acknowledgments}


%

\end{document}